\documentclass[11pt]{article}
\usepackage{amssymb,amsmath,amsfonts}
\usepackage{graphicx}
\usepackage{graphics}
\usepackage{eepic,epsfig}

\@addtoreset{equation}{section}

\makeatother

\begin{document}

\title{Finite temperature current densities and Bose-Einstein \\
condensation in topologically nontrivial spaces }
\author{E. R. Bezerra de Mello$^{1}$\thanks{%
E-mail: emello@fisica.ufpb.br}, \thinspace\ A. A. Saharian$^{1,2}$\thanks{%
E-mail: saharian@ysu.am} \vspace{0.3cm} \\
%EndAName
\textit{$^{1}$Departamento de F\'{\i}sica, Universidade Federal da Para\'{\i}%
ba}\\
\textit{58.059-970, Caixa Postal 5.008, Jo\~{a}o Pessoa, PB, Brazil}\vspace{%
0.3cm}\\
\textit{$^2$Department of Physics, Yerevan State University,}\\
\textit{1 Alex Manoogian Street, 0025 Yerevan, Armenia}}
\date{}
\maketitle

\begin{abstract}
We investigate the finite temperature expectation values of the charge and
current densities for a complex scalar field with nonzero chemical potential
in background of a flat spacetime with spatial topology $R^{p}\times
(S^{1})^{q}$. Along compact dimensions quasiperiodicity conditions with
general phases are imposed on the field. In addition, we assume the presence
of a constant gauge field which, due to the nontrivial topology of
background space, leads to Aharonov-Bohm-like effects on the expectation
values. By using the Abel-Plana-type summation formula and zeta function
techniques, two different representations are provided for both the current
and charge densities. The current density has nonzero components along the
compact dimensions only and, in the absence of a gauge field, it vanishes
for special cases of twisted and untwisted scalar fields. In the
high-temperature limit, the current density and the topological part in the
charge density are linear functions of the temperature. The Bose-Einstein
condensation for a fixed value of the charge is discussed. The expression
for the chemical potential is given in terms of the lengths of compact
dimensions, temperature and gauge field. It is shown that the parameters of
the phase transition can be controlled by tuning the gauge field. The
separate contributions to the charge and current densities coming from the
Bose-Einstein condensate and from excited states are also investigated.
\end{abstract}

\bigskip

PACS numbers: 03.70.+k, 11.10.Kk, 03.75.Hh

\bigskip

\section{Introduction}

In recent years, there has been a large interest to the physical problems
with compact spatial dimensions. Several models of this sort appear in high
energy physics, in cosmology and in condensed matter physics. In particular,
many of high energy theories of fundamental physics, including supergravity
and superstring theories, are formulated in spacetimes having extra compact
dimensions which are characterized by extremely small length scales. These
theories provide an attractive framework for the unification of
gravitational and gauge interactions. The models of a compact universe with
nontrivial topology may also play an important role by providing proper
initial conditions for inflation \cite{Lind04}.

In the models with compact dimensions, the nontrivial topology of background
space can have important physical implications in classical and quantum
field theories, which include instabilities in interacting field theories
\cite{Ford80a}, topological mass generation \cite{Ford79,Toms80b} and
symmetry breaking \cite{Toms80b,Odin88}. The periodicity conditions imposed
on fields along compact dimensions allow only the normal modes with suitable
wavelengths. As a result of this, the expectation values of various physical
observables are modified. In particular, many authors have investigated the
effects of vacuum or Casimir energies and stresses associated with the
presence of compact dimensions (for reviews see Refs. \cite{Most97}, \cite%
{Duff86}). The topological Casimir effect is a physical example of the
connection between quantum phenomena and global properties of spacetime. The
Casimir energy of bulk fields induces a non-trivial potential for the
compactification radius of higher-dimensional field theories providing a
stabilization mechanism for the corresponding moduli fields and thereby
fixing the effective gauge couplings. The Casimir effect has also been
considered as a possible origin for the dark energy in both Kaluza-Klein
type models and in braneworld scenario \cite{DarkEn}.

The main part of the papers, devoted to the influence of the nontrivial
topology on the properties of the quantum vacuum, considers the vacuum
energy and stresses. These quantities are chosen because of their close
connection with the structure of spacetime through the theory of
gravitation. For charged fields another important characteristic, bilinear
in the field, is the expectation value of the current density in a given
state. In Ref.~\cite{Bell10}, we have investigated the vacuum expectation
value of the current density for a fermionic field in spaces with an
arbitrary number of toroidally compactified dimensions. Application of the
general results are given to the electrons of a graphene sheet rolled into
cylindrical and toroidal shapes. For the description of the relevant
low-energy degrees of freedom we have used the effective field theory
treatment of graphene in terms of a pair of Dirac fermions. For this model
one has the topologies $R^{1}\times S^{1}$ and $(S^{1})^{2}$ for cylindrical
and toroidal nanotubes respectively. Combined effects of compact spatial
dimensions and boundaries on the vacuum expectation values of the fermionic
current have been discussed recently in Ref. \cite{Bell12}. In the latter,
the geometry of boundaries is given by two parallel plates on which the
fermion field obeys bag boundary conditions. The effects of nontrivial
topology around a conical defect on the current induced by a magnetic flux
were investigated in Ref. \cite{Srir01} for scalar and fermion fields.

In the present paper we consider the finite temperature charge and current
densities for a scalar field in background spacetime with spatial topology $%
R^{p}\times (S^{1})^{q}$. In both types of models with compact dimensions
used in the cosmology of the early Universe and in condensed matter physics,
the effects induced by the finite temperature play an important role. The
thermal corrections arise from thermal excitations of fluctuation spectrum
and they depend strongly on the geometry. As a consequence of this, thermal
modifications of quantum topological effects can differ qualitatively for
different geometries. The thermal Casimir effect in cosmological models with
nontrivial topology has been considered in Refs. \cite{Alta78}. A general
discussion of the finite temperature effects for a scalar field in higher
dimensional product manifolds with compact subspaces is given in Ref. \cite%
{Dowk84}. Specific calculations are presented for the cases when the
internal space is a torus or a sphere. In Ref. \cite{Campo91}, the
corresponding results are extended to the case in which a chemical potential
is present. In the previous discussions of the effects from nontrivial
topology and finite temperature, the authors mainly consider periodicity and
antiperiodicity conditions imposed on the field along compact dimensions.
The latter correspond to untwisted and twisted configurations of fields
respectively. In this case the current density corresponding to a conserved
charge associated with an internal symmetry vanishes. As it will be seen
below, the presence of a constant gauge field, interacting with a charged
quantum field, will induce a nontrivial phase in the periodicity conditions
along compact dimensions. As a consequence of this, nonzero components of
the current density appear along compact dimensions. This is a sort of
Aharonov-Bohm-like effect related to the nontrivial topology of the
background space.

The organization of the paper is as follows. In the next section the
geometry of the problem is described and the thermal Hadamard function is
evaluated for a complex scalar field in thermal equilibrium. In Section \ref%
{Sec:Charge}, by using the expression for the Hadamard function, the
expectation values of the charge and current densities are investigated.
Various limiting cases are discussed. Alternative expressions for the charge
and current densities are provided in Section \ref{Sec:Zeta} by making use
of the zeta function renormalization approach. The Section \ref{Sec:BEC} is
devoted to the investigation of the Bose-Einstein condensation in the
background under consideration. The properties of the vacuum expectation
value of the charge density are discussed in Appendix \ref{sec:Appendix}.
Throughout the paper we use the units $\hbar =c=k_{B}=1$, with $k_{B}$ been
Boltzmann constant.

\section{Geometry of the problem and the Hadamard function}

\label{Sec:Problem}

We consider the quantum scalar field $\varphi (x)$ on background of $\ (D+1)$
dimensional flat spacetime with spatial topology $R^{p}\times
(S^{1})^{q},p+q=D$. For the Cartesian coordinates along uncompactified and
compactified dimensions we use the notations $\mathbf{x}%
_{p}=(x^{1},...,x^{p})$ and $\mathbf{x}_{q}=(x^{p+1},...,x^{D})$,
respectively. The length of the $l-$th compact dimension we denote as $L_{l}$%
. Hence, for coordinates one has $-\infty <x^{l}<\infty $ for $l=1,..,p$,
and $0\leqslant x^{l}\leqslant L_{l}$ for $l=p+1,...,D$. In the presence of
a gauge field $A_{\mu }$ the field equation has the form%
\begin{equation}
\left( g^{\mu \nu }D_{\mu }D_{\nu }+m^{2}\right) \varphi =0,  \label{Feq}
\end{equation}%
where $D_{\mu }=\partial _{\mu }+ieA_{\mu }$ and $e$ is the charge
associated with the field. One of the characteristic features of field
theory on backgrounds with nontrivial topology is the appearance of
topologically inequivalent field configurations \cite{Isha78}. The boundary
conditions should be specified along the compact dimensions for the theory
to be defined. We assume that the field obeys generic quasiperiodic boundary
conditions,%
\begin{equation}
\varphi (t,\mathbf{x}_{p},\mathbf{x}_{q}+L_{l}\mathbf{e}_{l})=e^{i\alpha
_{l}}\varphi (t,\mathbf{x}_{p},\mathbf{x}_{q}),  \label{BC_pq}
\end{equation}%
with constant phases $|\alpha _{l}|\leqslant \pi $ and with $\mathbf{e}_{l}$
being the unit vector along the direction of the coordinate $%
x^{l},l=p+1,...,D$. The condition (\ref{BC_pq}) includes the periodicity
conditions for both untwisted and twisted scalar fields as special cases
with $\alpha _{l}=0$ and $\alpha _{l}=\pi $, respectively.

In the discussion below we will assume a constant gauge field $A_{\mu }$.
Though the corresponding field strength vanishes, the nontrivial topology of
the background spacetime leads to the Aharonov-Bohm-like effects on physical
observables. In the case of constant $A_{\mu }$, by making use of the gauge
transformation%
\begin{equation}
\varphi (x)=e^{-ie\chi }\varphi ^{\prime }(x),\;A_{\mu }=A_{\mu }^{\prime
}+\partial _{\mu }\chi ,  \label{gaugetrans}
\end{equation}%
with $\chi =A_{\mu }x^{\mu }$ we see that in the new gauge one has $A_{\mu
}^{\prime }=0$ and the vector potential disappears from the equation for $%
\varphi ^{\prime }(x)$. For the new field we have the periodicity condition%
\begin{equation}
\varphi ^{\prime }(t,\mathbf{x}_{p},\mathbf{x}_{q}+L_{l}\mathbf{e}_{l})=e^{i%
\tilde{\alpha}_{l}}\varphi ^{\prime }(t,\mathbf{x}_{p},\mathbf{x}_{q}),
\label{BCnew}
\end{equation}%
where%
\begin{equation}
\tilde{\alpha}_{l}=\alpha _{l}+eA_{l}L_{l}.  \label{alftilde}
\end{equation}%
In what follows we will work with the field $\varphi ^{\prime }(x)$ omitting
the prime. Note that for this field $D_{\mu }=\partial _{\mu }$. As it is
seen from Eq. (\ref{alftilde}), the presence of a constant gauge field
shifts the phases in the periodicity conditions along compact dimensions. In
particular, a nontrivial phase is induced for special cases of twisted and
untwisted fields. As it will be shown below, this is crucial for the
appearance of the nonzero current density along compact dimensions. Note
that the term in Eq. (\ref{alftilde}) due to the gauge field may be written
as%
\begin{equation}
eA_{l}L_{l}=2\pi \Phi _{l}/\Phi _{0},  \label{Flux}
\end{equation}%
where $\Phi _{l}$ is a formal flux enclosed by the circle corresponding to
the $l$-th compact dimension and $\Phi _{0}=2\pi /e$ is the flux quantum.

The complete set of positive- and negative-energy solutions for the problem
under consideration can be written in the form of plane waves:%
\begin{equation}
\varphi _{\mathbf{k}}^{(\pm )}(x)=C_{\mathbf{k}}e^{i\mathbf{k}\cdot \mathbf{r%
}\mp i\omega t},\;\omega _{\mathbf{k}}=\sqrt{\mathbf{k}^{2}+m^{2}},
\label{fisigma}
\end{equation}%
where $\mathbf{k}=(\mathbf{k}_{p},\mathbf{k}_{q})$, $\mathbf{k}%
_{p}=(k_{1},\ldots ,k_{p})$, $\mathbf{k}_{q}=(k_{p+1},\ldots ,k_{D})$, with $%
-\infty <k_{i}<+\infty $ for $i=1,\ldots ,p$. For the momentum components
along the compact dimensions the eigenvalues are determined from the
conditions (\ref{BCnew}):%
\begin{equation}
k_{l}=\left( 2\pi n_{l}+\tilde{\alpha}_{l}\right) /L_{l},\quad n_{l}=0,\pm
1,\pm 2,\ldots .,  \label{kltild}
\end{equation}%
with $l=p+1,...,D$. From Eq. (\ref{kltild}) it follows that the physical
results will depend on the fractional part of $\tilde{\alpha}_{l}/(2\pi )$
only. The integer part can be absorbed by the redefinition of $n_{l}$.
Hence, without loss of generality, we can assume that $|\tilde{\alpha}%
_{l}|\leqslant \pi $. The normalization coefficient in (\ref{fisigma}) is
found from the orthonormalization condition
\begin{equation}
\int d^{D}x\varphi _{\mathbf{k}}^{(\lambda )}(x)\varphi _{\mathbf{k}^{\prime
}}^{(\lambda ^{\prime })\ast }(x)=\frac{1}{2\omega _{\mathbf{k}}}\delta
_{\lambda \lambda ^{\prime }}\delta _{\mathbf{kk}^{\prime }},  \label{Norm}
\end{equation}%
where $\delta _{\mathbf{kk}^{\prime }}=\delta (\mathbf{k}_{p}-\mathbf{k}%
_{p}^{\prime })\delta _{n_{p+1},n_{p+1}^{\prime }}....\delta
_{n_{D},n_{D}^{\prime }}$. Substituting the functions (\ref{fisigma}), for
the normalization coefficient we find
\begin{equation}
|C_{\mathbf{k}}|^{2}=\frac{1}{2(2\pi )^{p}V_{q}\omega _{\mathbf{k}}},
\label{Cnorm}
\end{equation}%
with $V_{q}=L_{p+1}....L_{D}$ being the volume of the compact subspace and%
\begin{equation}
\omega _{\mathbf{k}}=\sqrt{\mathbf{k}_{p}^{2}+\mathbf{k}_{q}^{2}+m^{2}},\;%
\mathbf{k}_{q}^{2}=\sum_{l=p+1}^{D}\left( \frac{2\pi n_{l}+\tilde{\alpha}_{l}%
}{L_{l}}\right) ^{2}.  \label{omk}
\end{equation}%
The smallest value for the energy we will denote by $\omega _{0}$. Assuming
that $|\tilde{\alpha}_{l}|\leqslant \pi $, we have%
\begin{equation}
\omega _{0}=\sqrt{\sum\nolimits_{l=p+1}^{D}\tilde{\alpha}%
_{l}^{2}/L_{l}^{2}+m^{2}}.  \label{om0}
\end{equation}

We are interested in the expectation values of the charge and current
densities for the field $\varphi (x)$ in thermal equilibrium at finite
temperature $T$. These quantities can be evaluated by using the thermal
Hadamard function%
\begin{eqnarray}
G^{(1)}(x,x^{\prime }) &=&\left\langle \varphi (x)\varphi ^{+}(x^{\prime
})+\varphi ^{+}(x^{\prime })\varphi (x)\right\rangle  \notag \\
&=&\mathrm{tr}[\hat{\rho}(\varphi (x)\varphi ^{+}(x^{\prime })+\varphi
^{+}(x^{\prime })\varphi (x))],  \label{Wt}
\end{eqnarray}%
where $\left\langle \cdots \right\rangle $ means the ensemble average and $%
\hat{\rho}$ is the density matrix. For the thermodynamical equilibrium
distribution at temperature $T$, the latter is given by
\begin{equation}
\hat{\rho}=Z^{-1}e^{-\beta (\hat{H}-\mu ^{\prime }\hat{Q})},  \label{rho}
\end{equation}%
where $\beta =1/T$. In Eq. (\ref{rho}), $\widehat{Q}$ denotes a conserved
charge, $\mu ^{\prime }$ is the related chemical potential and $Z$ is the
grand-canonical partition function%
\begin{equation}
Z=\mathrm{tr}[e^{-\beta (\hat{H}-\mu ^{\prime }\hat{Q})}].  \label{PartFunc}
\end{equation}

In order to evaluate the expectation value in Eq. (\ref{Wt}) we expand the
field operator over a complete set of solutions:%
\begin{equation}
\varphi (x)=\sum_{\mathbf{k}}[\hat{a}_{\mathbf{k}}\varphi _{\mathbf{k}%
}^{(+)}(x)+\hat{b}_{\mathbf{k}}^{+}\varphi _{\mathbf{k}}^{(-)}(x)],
\label{phiexpand}
\end{equation}%
with $\sum_{\mathbf{k}}=\int d\mathbf{k}_{p}\sum_{\mathbf{n}_{q}}$ and $%
\mathbf{n}_{q}=(n_{p+1},\ldots ,n_{D})$. Here and in what follows we use the
notation
\begin{equation}
\sum_{\mathbf{n}}=\sum_{n_{1}=-\infty }^{+\infty }\cdots \sum_{n_{l}=-\infty
}^{+\infty },  \label{Not}
\end{equation}%
for $\mathbf{n}=(n_{1},\ldots ,n_{l})$. Substituting the expansion (\ref%
{phiexpand}) into Eq. (\ref{Wt}), we use the relations%
\begin{equation}
\text{\textrm{tr}}[\widehat{\rho }\hat{a}_{\mathbf{k}}^{+}\hat{a}_{\mathbf{k}%
^{\prime }}]=\frac{\delta _{\mathbf{kk}^{\prime }}}{e^{\beta (\omega _{%
\mathbf{k}}-\mu )}-1},\;\text{\textrm{tr}}[\widehat{\rho }\hat{b}_{\mathbf{k}%
}^{+}\hat{b}_{\mathbf{k}^{\prime }}]=\frac{\delta _{\mathbf{kk}^{\prime }}}{%
e^{\beta (\omega _{\mathbf{k}}+\mu )}-1},  \label{RelTr}
\end{equation}%
where $\mu =e\mu ^{\prime }$. Note that the chemical potentials have
opposite signs for particles ($\mu $) and antiparticles ($-\mu $). The
expectation values for $\hat{a}_{\mathbf{k}}\hat{a}_{\mathbf{k}^{\prime
}}^{+}$ and $\hat{b}_{\mathbf{k}}\hat{b}_{\mathbf{k}^{\prime }}^{+}$ are
obtained from (\ref{RelTr}) by using the commutation relations and the
expectation values for the other products are zero. For the Hadamard
function we get%
\begin{equation}
G^{(1)}(x,x^{\prime })=G_{0}^{(1)}(x,x^{\prime })+2\sum_{\mathbf{k}%
}\sum_{s=\pm }\frac{\varphi _{\mathbf{k}}^{(s)}(x)\varphi _{\mathbf{k}%
}^{(s)\ast }(x^{\prime })}{e^{\beta (\omega _{\mathbf{k}}-s\mu )}-1},
\label{WF1}
\end{equation}%
where the first term in the right-hand side corresponds to the zero
temperature Hadamard function:%
\begin{eqnarray}
G_{0}^{(1)}(x,x^{\prime }) &=&\left\langle 0\right\vert \varphi (x)\varphi
^{+}(x^{\prime })+\varphi ^{+}(x^{\prime })\varphi (x)\left\vert
0\right\rangle  \notag \\
&=&\sum_{\mathbf{k}}\sum_{s=\pm }\varphi _{\mathbf{k}}^{(s)}(x)\varphi _{%
\mathbf{k}}^{(s)\ast }(x^{\prime }),  \label{G01}
\end{eqnarray}%
with $\left\vert 0\right\rangle $ being the vacuum state. In order to ensure
a positive-definite value for the number of particles we assume that $|\mu
|\leqslant \omega _{0}$, where $\omega _{0}$ is the smallest value of the
energy (see Eq. (\ref{om0})).

By using the expressions (\ref{fisigma}) for the mode functions and the
expansion $(e^{y}-1)^{-1}=\sum_{n=1}^{\infty }e^{-ny}$, the mode sum for the
Hadamard function is written in the form%
\begin{eqnarray}
&&G^{(1)}(x,x^{\prime })=\frac{1}{V_{q}}\int \frac{d\mathbf{k}_{p}}{(2\pi
)^{p}}e^{i\mathbf{k}_{p}\cdot \Delta \mathbf{x}_{p}}\sum_{\mathbf{n}_{q}}%
\frac{e^{i\mathbf{k}_{q}\cdot \Delta \mathbf{x}_{q}}}{\omega _{\mathbf{k}}}
\notag \\
&&\qquad \times \left[ \cos (\omega _{\mathbf{k}}\Delta
t)+\sum_{n=1}^{\infty }\sum_{s=\pm }e^{\omega _{\mathbf{k}}\left( si\Delta
t-n\beta \right) -sn\mu \beta }\right] ,  \label{WF2}
\end{eqnarray}%
where $\Delta \mathbf{x}_{p}\mathbf{=x}_{p}-\mathbf{x}_{p}^{\prime }$, $%
\Delta \mathbf{x}_{q}\mathbf{=x}_{q}-\mathbf{x}_{q}^{\prime }$, $\Delta
t=t-t^{\prime }$. For the evaluation of the Hadamard function we apply to
the series over $n_{r}$ the Abel-Plana-type summation formula \cite%
{Beze08,Bell09} (for applications of the Abel-Plana formula and its
generalizations in quantum field theory see Refs. \cite{Most97,Mama76,Saha08}%
)%
\begin{eqnarray}
&&\frac{2\pi }{L_{r}}\sum_{n_{r}=-\infty }^{\infty }g(k_{r})f(\left\vert
k_{r}\right\vert )=\int_{0}^{\infty }dz[g(z)+g(-z)]f(z)  \notag \\
&&\qquad +i\int_{0}^{\infty }dz\,[f(iz)-f(-iz)]\sum_{\lambda =\pm 1}\frac{%
g(i\lambda z)}{e^{zL_{r}+i\lambda \tilde{\alpha}_{r}}-1},  \label{AbelPlan1}
\end{eqnarray}%
where $k_{r}$ is given by Eq. (\ref{kltild}). For the Hadamard function we
find the expression%
\begin{eqnarray}
&&G^{(1)}(x,x^{\prime })=G_{p+1,q-1}^{(1)}(x,x^{\prime })+\frac{L_{r}}{\pi
V_{q}}\int \frac{d\mathbf{k}_{p}}{(2\pi )^{p}}\sum_{\mathbf{n}_{q-1}^{r}}e^{i%
\mathbf{k}_{p}\cdot \Delta \mathbf{x}_{p}+i\mathbf{k}_{q-1}\cdot \Delta
\mathbf{x}_{q-1}}  \notag \\
&&\qquad \times \sum_{n=-\infty }^{\infty }e^{\mu n\beta }\int_{\omega
_{p,q-1}}^{\infty }dz\frac{\cosh [(\Delta t-in\beta )\sqrt{z^{2}-\omega
_{p,q-1}^{2}}]}{\sqrt{z^{2}-\omega _{p,q-1}^{2}}}\sum_{\lambda =\pm 1}\frac{%
e^{-\lambda z\Delta x^{r}}}{e^{zL_{r}+\lambda i\tilde{\alpha}_{r}}-1},
\label{WF3}
\end{eqnarray}%
where $\mathbf{n}_{q-1}^{r}=(n_{p+1},\ldots ,n_{r-1},n_{r+1},\ldots ,n_{D})$%
, $\mathbf{k}_{q-1}=(k_{p+1},\ldots ,k_{r-1},k_{r+1},\ldots ,k_{D})$, and%
\begin{equation}
\omega _{p,q-1}=\sqrt{\mathbf{k}_{p}^{2}+\mathbf{k}_{q-1}^{2}+m^{2}}.
\label{ompq-1}
\end{equation}%
The first term in the right-hand side of Eq. (\ref{WF3}), $%
G_{p+1,q-1}^{(1)}(x,x^{\prime })$, comes from the first term on the right of
Eq. (\ref{AbelPlan1}) and it is the Hadamard function for the topology $%
R^{p+1}\times (S^{1})^{q-1}$ with the lengths of the compact dimensions $%
(L_{p+1},\ldots ,L_{r-1},L_{r+1},\ldots ,L_{D})$.

For the further transformation of the expression (\ref{WF3}) we use the
expansion%
\begin{equation}
\frac{e^{-\lambda z\Delta x^{r}}}{e^{zL_{r}+\lambda i\tilde{\alpha}_{r}}-1}%
=\sum_{l=1}^{\infty }e^{-z(lL_{r}+\lambda \Delta x^{r})-\lambda il\tilde{%
\alpha}_{r}}.  \label{exp1}
\end{equation}%
With this expansion the $z$-integral is expressed in terms of the Macdonald
function of the zeroth order. Then the integral over $\mathbf{k}_{p}$ is
evaluated by using the formula from Ref. \cite{Prud86}. For the Hadamard
function we arrive to the final expression%
\begin{eqnarray}
&& G^{(1)}(x,x^{\prime }) =\frac{2L_{r}V_{q}^{-1}}{(2\pi )^{p/2+1}}%
\sum_{n=-\infty }^{\infty }\sum_{\mathbf{n}_{q}}e^{in_{r}\tilde{\alpha}%
_{r}+n\mu \beta }e^{i\mathbf{k}_{q-1}\cdot \Delta \mathbf{x}_{q-1}}  \notag
\\
&& \qquad \times \omega _{\mathbf{n}_{q-1}^{r}}^{p}f_{p/2}\left( \omega _{%
\mathbf{n}_{q-1}^{r}}\sqrt{|\Delta \mathbf{x}_{p}|^{2}+\left( \Delta
x^{r}-n_{r}L_{r}\right) ^{2}-\left( \Delta t-in\beta \right) ^{2}}\right) ,
\label{WF4}
\end{eqnarray}%
where%
\begin{equation}
f_{\nu }(x)=x^{-\nu }K_{\nu }(x),\;\omega _{\mathbf{n}_{q-1}^{r}}=\sqrt{%
\mathbf{k}_{q-1}^{2}+m^{2}}.  \label{fnu}
\end{equation}%
Note that the $n_{r}=0$ term in Eq. (\ref{WF4}) corresponds to the function $%
G_{p+1,q-1}^{(1)}(x,x^{\prime })$. Hence, the part of the Hadamard function
in Eq. (\ref{WF4}) with $n_{r}\neq 0$ is induced by the compactification of
the $r$-th direction to a circle with the length $L_{r}$.

An alternative expression for the Hadamard function is obtained directly
from Eq. (\ref{WF2}). We first integrate over the angular part of $\mathbf{k}%
_{p}$ and then the integral over $|\mathbf{k}_{p}|$ is expressed in terms of
the Macdonald function. The corresponding expression is written in terms of
the function (\ref{fnu}) as%
\begin{eqnarray}
G^{(1)}(x,x^{\prime }) &=&\frac{2V_{q}^{-1}}{(2\pi )^{\frac{p+1}{2}}}\sum_{%
\mathbf{n}_{q}}e^{i\mathbf{k}_{q}\cdot \Delta \mathbf{x}_{q}}\omega _{%
\mathbf{n}_{q}}^{p-1}\sum_{n=-\infty }^{+\infty }e^{n\mu \beta }  \notag \\
&&\times f_{\frac{p-1}{2}}(\omega _{\mathbf{n}_{q}}\sqrt{|\Delta \mathbf{x}%
_{p}|^{2}-(\Delta t-in\beta )^{2}}),  \label{G1alt}
\end{eqnarray}%
with the notation
\begin{equation}
\omega _{\mathbf{n}_{q}}=\sqrt{\mathbf{k}_{q}^{2}+m^{2}},  \label{omnq1}
\end{equation}%
and $\mathbf{k}_{q}^{2}$ is given by Eq. (\ref{omk}). Note that the explicit
information contained in Eq. (\ref{WF4}) is more detailed. Both
representations (\ref{WF4}) and (\ref{G1alt}) present the thermal Hadamard
function as an infinite imaginary-time image sum of the zero temperature
Hadamard function. This is the well-known result in finite temperature field
theory (see, for instance, Ref. \cite{Birr82}).

\section{Charge density}

\label{Sec:Charge}

Having the thermal Hadamard function we can evaluate the expectation value
for the current density%
\begin{equation}
j_{l}(x)=ie[\varphi ^{+}(x)\partial _{l}\varphi (x)-(\partial _{l}\varphi
^{+}(x))\varphi (x)],  \label{jmu}
\end{equation}%
$l=0,1,\ldots ,D$, by using the formula%
\begin{equation}
\left\langle j_{l}(x)\right\rangle =\frac{i}{2}e\lim_{x^{\prime }\rightarrow
x}(\partial _{l}-\partial _{l}^{\prime })G^{(1)}(x,x^{\prime }).  \label{jW}
\end{equation}%
By making use of the relation $\partial _{z}f_{\nu }(z)=-zf_{\nu +1}(z)$,
from Eq. (\ref{WF4}) for the charge density ($l=0$) one finds%
\begin{eqnarray}
\left\langle j_{0}\right\rangle &=&\frac{8e\beta L_{r}}{(2\pi )^{\frac{p}{2}%
+1}V_{q}}\sideset{}{'}{\sum}_{n_{r}=0}^{\infty }\cos (n_{r}\tilde{\alpha}%
_{r})\sum_{n=1}^{\infty }n\sinh (n\mu \beta )  \notag \\
&&\times \sum_{\mathbf{n}_{q-1}^{r}}\omega _{\mathbf{n}_{q-1}^{r}}^{p+2}f_{%
\frac{p}{2}+1}(\omega _{\mathbf{n}_{q-1}^{r}}\sqrt{n_{r}^{2}L_{r}^{2}+n^{2}%
\beta ^{2}}),  \label{j0}
\end{eqnarray}%
where the prime on the sign of sum means that the term $n_{r}=0$ should be
taken with the coefficient 1/2.

As it is seen from Eq. (\ref{j0}), the charge density is an even function of
the phases $\tilde{\alpha}_{l}$ and, for a fixed value of the chemical
potential, it vanishes in the zero temperature limit. It is a periodic
function of $\tilde{\alpha}_{l}$ with the period equal to $2\pi $. In the
case of zero chemical potential the charge density is zero. In Eq. (\ref{j0}%
), the term with $n_{r}=0$ corresponds to the charge density for the
topology $R^{p+1}\times (S^{1})^{q-1}$ with the lengths of the compact
dimensions $(L_{p+1},\ldots ,L_{r-1},L_{r+1},\ldots ,L_{D})$ and the
contribution of the terms with $n_{r}\neq 0$ is the change in the charge
density due to the compactification of the $r$-th dimension to $S^{1}$ with
the length $L_{r}$. By taking into account Eq. (\ref{Flux}), we see that the
charge density is a periodic function of fluxes $\Phi _{l}$ with the period
equal to the flux quantum. Note that the sign of the ratio $\left\langle
j_{0}\right\rangle /e$ coincides with the sign of the chemical potential.

An alternative expression for the charge density, more symmetric with
respect to the compact dimensions, is obtained by applying the formula%
\begin{equation}
\sum_{n=-\infty }^{+\infty }\cos (n\alpha )f_{\nu }(c\sqrt{b^{2}+a^{2}n^{2}}%
)=\frac{\sqrt{2\pi }}{ac^{2\nu }}\sum_{n=-\infty }^{+\infty }w_{n}^{2\nu
-1}f_{\nu -1/2}(bw_{n}),  \label{Rel3n}
\end{equation}%
with $a,b,c>0$, $w_{n}=\sqrt{(2\pi n+\alpha )^{2}/a^{2}+c^{2}}$, to the
series over $n_{r}$ in Eq. (\ref{j0}). This leads to the expression%
\begin{equation}
\left\langle j_{0}\right\rangle =\frac{4e\beta V_{q}^{-1}}{(2\pi )^{\frac{p+1%
}{2}}}\sum_{n=1}^{\infty }n\sinh (n\mu \beta )\sum_{\mathbf{n}_{q}}\omega _{%
\mathbf{n}_{q}}^{p+1}f_{\frac{p+1}{2}}(n\beta \omega _{\mathbf{n}_{q}}),
\label{j01}
\end{equation}%
with the notation (\ref{omnq1}). This formula could also be directly
obtained from Eq. (\ref{jW}) using the expression (\ref{G1alt}) for the
Hadamard function. The form (\ref{j01}) for the charge density in the case
of topology $R^{p+1}\times (S^{1})^{q-1}$ is also obtained from Eq. (\ref{j0}%
) taking the limit $L_{r}\rightarrow \infty $.

In the case of Minkowski spacetime one has $p=D$, $q=0$, and from Eq. (\ref%
{j01}) we get%
\begin{equation}
\left\langle j_{0}\right\rangle _{\mathrm{(M)}}=\frac{4e\beta m^{D+1}}{(2\pi
)^{\frac{D+1}{2}}}\sum_{n=1}^{\infty }n\sinh (n\mu \beta )f_{\frac{D+1}{2}%
}(n\beta m),  \label{j01M}
\end{equation}%
with $|\mu |\leqslant m$. The thermodynamic properties of the relativistic
Bose gas in this case have been considered in Refs. \cite{Kapu81,Habe81}. If
all spatial dimensions are compactified, the corresponding formulas are
obtained from Eqs. (\ref{j0}) and (\ref{j01}) taking $p=0$. In particular,
from Eq. (\ref{j01}) one has%
\begin{equation}
\left\langle j_{0}\right\rangle =\frac{2e}{V_{q}}\sum_{n=1}^{\infty }\sinh
(n\mu \beta )\sum_{\mathbf{n}_{q}}e^{-n\beta \omega _{\mathbf{n}_{q}}},
\label{j0p0}
\end{equation}%
where we have used $f_{1/2}(x)=\sqrt{\pi /2}x^{-1}e^{-x}$.

Let us consider some limiting cases of Eq. (\ref{j01}). If the length of the
$l$-th compact dimensions is large compared to other length scales, in the
sum over $n_{l}$ in Eq. (\ref{j01}) the contribution from large values of $%
n_{l}$ dominates and, to the leading order, we replace the summation by the
integration. The corresponding integral is evaluated with the help of the
formula
\begin{equation}
\int_{0}^{\infty }dy(y^{2}+b^{2})^{\frac{p+1}{2}}f_{\frac{p+1}{2}}(c\sqrt{%
y^{2}+b^{2}})=\sqrt{\frac{\pi }{2}}b^{p+2}f_{\frac{p}{2}+1}(cb),
\label{IntForm3}
\end{equation}%
and from Eq. (\ref{j01}) we obtain the expression of the charge density for
the topology $R^{p+1}\times (S^{1})^{q-1}$.

If the length of the $l$-th compact dimension is small compared with the
other length scales and $L_{l}\ll \beta $, under the assumption $|\tilde{%
\alpha}_{l}|<\pi $, the main contribution to the corresponding series in Eq.
(\ref{j01}) comes from the term with $n_{l}=0$. The behavior of the charge
density is essentially different with dependence whether the phase $\tilde{%
\alpha}_{l}$ is zero or not. When $\tilde{\alpha}_{l}=0$, we can see that,
to the leading order, $L_{l}\left\langle j_{0}\right\rangle $ coincides with
the charge density in $(D-1)$-dimensional space of topology $R^{p}\times
(S^{1})^{q-1}$ and with the lengths of the compact dimensions $L_{p+1}$%
,\ldots ,$L_{l-1}$,$L_{l+1}$,\ldots ,$L_{D}$. In particular, this is the
case for an untwisted scalar field in the absence of a gauge field. For $%
\tilde{\alpha}_{l}\neq 0$ and for small values of $L_{l}$, the argument of
the Macdonald function in Eq. (\ref{j01}) is large and the charge density is
suppressed by the factor $e^{-|\tilde{\alpha}_{l}|\beta /L_{l}}$.

In the low-temperature limit the parameter $\beta $ is large and the
dominant contribution to the charge density comes from the term $n=1$ in the
series over $n$ and from the term in the series over $\mathbf{n}_{q}$ with
the smallest value of $\omega _{\mathbf{n}_{q}}$ which corresponds to $%
n_{l}=0$, $l=p+1,\ldots ,D$. To the leading order we find%
\begin{equation}
\left\langle j_{0}\right\rangle \approx \frac{4eV_{q}^{-1}\mathrm{sgn}(\mu )%
}{(2\pi )^{p/2+1}\beta ^{p/2}}\omega _{0}^{p/2}e^{-\beta \omega _{0}+|\mu
|\beta },  \label{j02lowT}
\end{equation}%
with $\omega _{0}$ given by Eq. (\ref{om0}).

From Eq. (\ref{j01}) it follows that the expectation value of the charge
density is finite in the limit $|\mu |\rightarrow \omega _{0}$ for $p>2$ and
it diverges for $p\leqslant 2$. In order to find the asymptotic behavior
near the point $|\mu |=\omega _{0}$, we note that for $p\leqslant 2$, under
the condition $\beta (\omega _{0}-|\mu |)\ll 1$, the main contribution to
Eq. (\ref{j01}) comes from the term with $n_{l}=0$ ($\omega _{\mathbf{n}%
_{q}}=\omega _{0}$) and in the corresponding series over $n$ the
contribution of large $n$ dominates. In this case we can use the asymptotic
expression for the Macdonald function for large values of the argument and
to the leading order this gives:%
\begin{equation}
\left\langle j_{0}\right\rangle \approx \mathrm{sgn}(\mu )\frac{e}{V_{q}}%
\left( \frac{\omega _{0}}{2\pi \beta }\right) ^{p/2}\mathrm{Li}%
_{p/2}(e^{-\beta (\omega _{0}-|\mu |)}),  \label{muom}
\end{equation}%
where $\mathrm{Li}_{s}(x)$ is the polylogarithm function. For the latter one
has $\mathrm{Li}_{0}(x)=x/(1-x)$, $\mathrm{Li}_{1}(x)=-\ln (1-x)$. By taking
into account that $\mathrm{Li}_{s}(e^{-y})\approx \Gamma (1-s)y^{s-1}$ for $%
|y|\ll 1$ and $s<1$, one finds the following asymptotic expressions:%
\begin{eqnarray}
\left\langle j_{0}\right\rangle &\approx &eT\frac{\mathrm{sgn}(\mu )\Gamma
(1-p/2)}{V_{q}(\omega _{0}-|\mu |)^{1-p/2}}\left( \frac{\omega _{0}}{2\pi }%
\right) ^{p/2},\;p=0,1,  \notag \\
\left\langle j_{0}\right\rangle &\approx &-eT\frac{\omega _{0}\mathrm{sgn}%
(\mu )}{2\pi V_{q}}\ln [(\omega _{0}-|\mu |)/T],\;p=2.  \label{j0mutoom0}
\end{eqnarray}

\begin{figure}[tbph]
\begin{center}
\begin{tabular}{cc}
\epsfig{figure=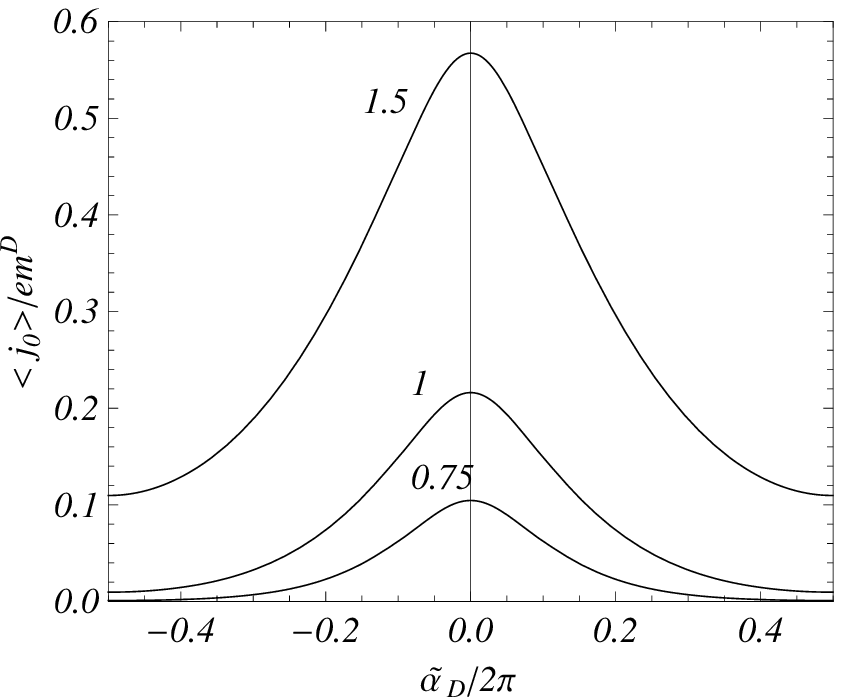,width=7.cm,height=6.cm} & \quad %
\epsfig{figure=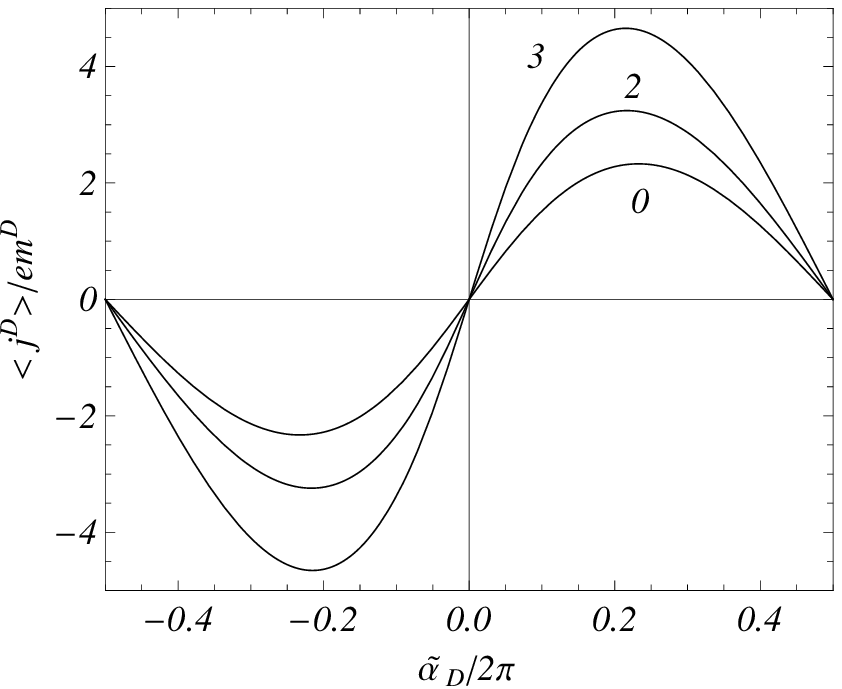,width=7.cm,height=6.cm}%
\end{tabular}%
\end{center}
\caption{The expectation values of the charge (left plot) and current (right
plot) densities as functions of the parameter $\tilde{\protect\alpha}_{D}/2%
\protect\pi $ for the $D=4$ model with a single compact dimension and for $%
\protect\mu =0.5m$, $mL_{D}=0.5$. The numbers near the curves correspond to
the values of $T/m$.}
\label{fig1}
\end{figure}

In the left plot of figure \ref{fig1} we present the charge density as a
function of the parameter $\tilde{\alpha}_{D}/(2\pi )$ in the $D=4$ model
with a single compact dimension of the length $L_{D}$. Note that for an
untwisted scalar field this parameter is the flux measured in units of the
flux quantum. For the chemical potential and for the length of the compact
dimensions we have taken the values corresponding to $\mu =0.5m$ and $%
mL_{D}=0.5$. The numbers near the curves correspond to the values of $T/m$.

\section{Current density}

\label{sec:Current}

Now we turn to the expectation value of the current density. As it can be
easily seen, the components of the current density along the uncompactified
dimensions vanish: $\left\langle j_{r}\right\rangle =0$ for $r=1,\ldots ,p$.
By making use of Eq. (\ref{jW}) and the expression (\ref{WF4}) of the
Hadamard function, for the current density along the $r$-th compact
dimension we get:%
\begin{eqnarray}
\left\langle j^{r}\right\rangle &=&\frac{8eL_{r}^{2}V_{q}^{-1}}{(2\pi
)^{p/2+1}}\sideset{}{'}{\sum}_{n=0}^{\infty }\cosh (\mu n\beta
)\sum_{n_{r}=1}^{\infty }n_{r}\sin (n_{r}\tilde{\alpha}_{r})  \notag \\
&&\times \sum_{\mathbf{n}_{q-1}^{r}}\omega _{\mathbf{n}_{q-1}^{r}}^{p+2}f_{%
\frac{p}{2}+1}(\omega _{\mathbf{n}_{q-1}^{r}}\sqrt{n_{r}^{2}L_{r}^{2}+n^{2}%
\beta ^{2}}),  \label{jr}
\end{eqnarray}%
with $r=p+1,\ldots ,D$ and, as before, the prime means that the term with $%
n=0$ should be taken with the weight 1/2. Note that, unlike to the case of
the charge density, the current density does not vanish at zero temperature
for a fixed value of the chemical potential. The zero temperature current
density is given by the $n=0$ term in Eq. (\ref{jr}):%
\begin{equation}
\left\langle j^{r}\right\rangle _{0}=\frac{4eL_{r}^{2}V_{q}^{-1}}{(2\pi
)^{p/2+1}}\sum_{n_{r}=1}^{\infty }n_{r}\sin (n_{r}\tilde{\alpha}_{r})\sum_{%
\mathbf{n}_{q-1}^{r}}\omega _{\mathbf{n}%
_{q-1}^{r}}^{p+2}f_{p/2+1}(n_{r}L_{r}\omega _{\mathbf{n}_{q-1}^{r}}).
\label{jrT0}
\end{equation}%
The features of this current are discussed in detail in Appendix \ref%
{sec:Appendix}. For the model with a single compact dimension the general
formula reduces to:%
\begin{eqnarray}
\left\langle j^{r}\right\rangle &=&\frac{8eL_{r}m^{D+1}}{(2\pi )^{\frac{D+1}{%
2}}}\sideset{}{'}{\sum}_{n=0}^{\infty }\cosh (n\mu \beta
)\sum_{n_{r}=1}^{\infty }n_{r}  \notag \\
&&\times \sin (n_{r}\tilde{\alpha}_{r})f_{\frac{D+1}{2}}(m\sqrt{%
n_{r}^{2}L_{r}^{2}+n^{2}\beta ^{2}}).  \label{jrsing}
\end{eqnarray}

An alternative expression of the current density is obtained by making use
of the formula (\ref{G1alt}) for the Hadamard function in Eq. (\ref{jW}):%
\begin{eqnarray}
\left\langle j^{r}\right\rangle &=&\left\langle j^{r}\right\rangle _{0}+%
\frac{4eV_{q}^{-1}}{(2\pi )^{\frac{p+1}{2}}L_{r}}\sum_{n=1}^{\infty }\cosh
(n\mu \beta )  \notag \\
&&\times \sum_{\mathbf{n}_{q}}\left( 2\pi n_{r}+\tilde{\alpha}_{r}\right)
\omega _{\mathbf{n}_{q}}^{p-1}f_{\frac{p-1}{2}}(n\beta \omega _{\mathbf{n}%
_{q}}).  \label{jr2}
\end{eqnarray}%
From Eqs. (\ref{jr}) and (\ref{jr2}) it follows that, the current density
along the $r$-th compact dimension is an odd periodic function of $\tilde{%
\alpha}_{r}$ and an even periodic function of $\tilde{\alpha}_{l}$, $l\neq r$%
, with the period equal to $2\pi $. The current density is an even function
of the chemical potential and it does not vanish in the limit of zero
chemical potential. In the absence of uncompactified dimensions one has $p=0$
and from Eq. (\ref{jr2}) we get%
\begin{equation}
\left\langle j^{r}\right\rangle =\left\langle j^{r}\right\rangle _{0}+\frac{%
2e}{V_{q}L_{r}}\sum_{n=1}^{\infty }\cosh (n\mu \beta )\sum_{\mathbf{n}%
_{q}}\left( 2\pi n_{r}+\tilde{\alpha}_{r}\right) \frac{e^{-n\beta \omega _{%
\mathbf{n}_{q}}}}{\omega _{\mathbf{n}_{q}}},  \label{jrp0}
\end{equation}%
where we have used $f_{-1/2}(x)=\sqrt{\pi /2}e^{-x}$. Here we assume that $%
\omega _{0}>0$. In the case $\omega _{0}=0$ there is a zero mode and the
contribution of this mode should be considered separately.

In a way similar to that for the case of the charge density, we can see that
in the limit when the length of the $l$-th compact dimension is large ($%
l\neq r$), the leading term obtained from Eq. (\ref{jr2}) coincides with the
current density in the space with topology $R^{p+1}\times (S^{1})^{q-1}$
with the lengths of the compact dimensions $L_{p+1}$,\ldots ,$L_{l-1}$,$%
L_{l+1}$,\ldots ,$L_{D}$. For small values of $L_{l}$, $l\neq r$, the
behavior of the current density crucially depends whether $\tilde{\alpha}%
_{l} $ is zero or not. For $\tilde{\alpha}_{l}=0$ the dominant contribution
comes from the term with $n_{l}=0$ and from the expression given above we
can see that, to the leading order, $L_{l}\left\langle j^{r}\right\rangle $
coincides with the corresponding quantity in $(D-1)$-dimensional space with
topology $R^{p}\times (S^{1})^{q-1}$ and with the lengths of the compact
dimensions $L_{p+1}$,\ldots ,$L_{l-1}$,$L_{l+1}$,\ldots ,$L_{D}$. For $%
\tilde{\alpha}_{l}\neq 0$ and for small values of $L_{l}$, the current
density $\left\langle j^{r}\right\rangle $ is exponentially suppressed.

If $L_{r}\gg \beta $, the dominant contribution to the series over $n$ in
Eq. (\ref{jrsing}) comes from large values of $n\sim L_{r}/\beta $. In this
case we can replace the summation by the integration and the corresponding
integral is evaluated by using the formula from Ref. \cite{Prud86} (assuming
that $|\mu |<\omega _{\mathbf{n}_{q-1}^{r}}$). To the leading order we get
\begin{eqnarray}
\left\langle j^{r}\right\rangle &\approx &\frac{4eL_{r}^{2}V_{q}^{-1}T}{%
(2\pi )^{(p+1)/2}}\sum_{l=1}^{\infty }n_{r}\sin (n_{r}\tilde{\alpha}%
_{r})\sum_{\mathbf{n}_{q-1}^{r}}(\omega _{\mathbf{n}_{q-1}^{r}}^{2}-\mu
^{2})^{\frac{p+1}{2}}  \notag \\
&&\times f_{(p+1)/2}(n_{r}L_{r}\sqrt{\omega _{\mathbf{n}_{q-1}^{r}}^{2}-\mu
^{2}}).  \label{jras}
\end{eqnarray}%
For a fixed value of $L_{r}$ this formula gives the leading term in the
high-temperature asymptotic for the current density. If in addition $%
L_{r}\gg L_{l}$, $l\neq r$, the dominant contribution comes from the term
with $n_{r}=1$, $n_{l}=0$, and the current density $\left\langle
j^{r}\right\rangle $ is suppressed by the factor $e^{-L_{r}\sqrt{\omega
_{0r}^{2}-\mu ^{2}}}$, where%
\begin{equation}
\omega _{0r}=\sqrt{\sum\nolimits_{l=p+1,\neq r}^{D}\tilde{\alpha}%
_{l}^{2}/L_{l}^{2}+m^{2}}.  \label{om0r}
\end{equation}

In order to see the asymptotic behavior of the current density at low
temperatures it is more convenient to use Eq. (\ref{jr2}). Assuming that $%
\beta \left( \omega _{0}-|\mu |\right) \gg 1$, the dominant contribution to
the temperature dependent part comes from the mode with the smallest energy
corresponding to $n_{l}=0$ and one has
\begin{equation}
\left\langle j^{r}\right\rangle \approx \left\langle j^{r}\right\rangle
_{0}+e\tilde{\alpha }_{r}\frac{\omega _{0}^{p/2-1}e^{-\beta \omega _{0}+|\mu
|\beta }}{(2\pi )^{p/2}V_{q}L_{r}\beta ^{p/2}}.  \label{jr2low}
\end{equation}%
In this case the temperature corrections are exponentially small.

For $p\leqslant 2$ the current density, defined by Eq. (\ref{jr2}), is
divergent in the limit $|\mu |\rightarrow \omega _{0}$. The corresponding
asymptotic is found in a way similar to that for the case of the charge
density. To the leading order we have%
\begin{equation}
\left\langle j^{r}\right\rangle \approx \frac{\tilde{\alpha}_{r}\mathrm{sgn}%
(\mu )}{L_{r}\omega _{0}}\left\langle j_{0}\right\rangle ,  \label{jrmutoom0}
\end{equation}%
where the asymptotic expressions for $\left\langle j_{0}\right\rangle $ for
separate values of $p$ are given in Eq. (\ref{j0mutoom0}).

In the right plot of figure \ref{fig1} we displayed the current density
along the compact dimension $x^{D}$ as a function of $\tilde{\alpha}%
_{D}/(2\pi )$ for the $D=4$ model with a single compact dimension of the
length corresponding to $mL_{D}=0.5$. The numbers near the curves are the
values of $T/m$ and for the chemical potential we have taken the value $\mu
=0.5m$.

\section{Zeta function approach}

\label{Sec:Zeta}

The expectation values of the charge and current densities can be evaluated
directly from Eq. (\ref{jmu}) by using zeta function techniques (see, for
instance, Ref. \cite{Kirs01}). First we consider the current density.

\subsection{Current density}

Substituting the expansion (\ref{phiexpand}) for the field operator and by
making use the expression (\ref{fisigma}) for the mode functions, for the
current density along compact dimensions one finds the following expression%
\begin{equation}
\left\langle j^{r}\right\rangle =\frac{e}{(2\pi )^{p}V_{q}}\sum_{\mathbf{k}}%
\frac{k_{r}}{\omega _{\mathbf{k}}}\left[ 1+\sum_{s=\pm }\frac{1}{e^{\beta
\left( \omega _{\mathbf{k}}-s\mu \right) }-1}\right] ,  \label{jrmode}
\end{equation}%
with $k_{r}=\left( 2\pi n_{r}+\tilde{\alpha}_{r}\right) /L_{r}$ and $%
r=p+1,\ldots ,D$. The first term in the square brackets corresponds to the
current density at zero temperature. The $s=+/-$ terms are contribution
coming from the particles/antiparticles. For the further transformations it
is convenient to write Eq. (\ref{jrmode}) in the form%
\begin{equation}
\left\langle j^{r}\right\rangle =\frac{2e}{(2\pi )^{p}V_{q}}\sum_{\mathbf{k}}%
\frac{k_{r}}{\omega _{\mathbf{k}}}\sideset{}{'}{\sum}_{n=0}^{\infty
}e^{-n\beta \omega _{\mathbf{k}}}\cosh (n\beta \mu ).  \label{jmu1}
\end{equation}%
In the special case $p=0$ this formula is reduced to Eq. (\ref{jrp0}). In
the representation (\ref{jmu1}), the zero temperature part corresponds to
the $n=0$ term. The divergences are contained in this part only. The
components of the current density along uncompact dimensions vanish.

As the next step, in Eq. (\ref{jmu1}) we use the integral representation%
\begin{equation}
\frac{e^{-n\beta \omega }}{\omega }=\frac{2}{\sqrt{\pi }}\int_{0}^{\infty
}ds\,e^{-\omega ^{2}s^{2}-n^{2}\beta ^{2}/4s^{2}}.  \label{IntRep1}
\end{equation}%
This allows us to write the expectation value of the current density in the
form
\begin{equation}
\left\langle j^{r}\right\rangle =\frac{2\pi ^{-1/2}e}{(2\pi )^{p}V_{q}}\sum_{%
\mathbf{k}}k_{r}\int_{0}^{\infty }ds\,e^{-\omega _{\mathbf{k}%
}^{2}s^{2}}\sum_{n=-\infty }^{\infty }e^{n\beta \mu -n^{2}\beta ^{2}/4s^{2}}.
\label{jrz}
\end{equation}%
Now we apply to the series over $n$ the Poisson summation formula%
\begin{equation}
\sum_{n=-\infty }^{+\infty }g(n\alpha )=\frac{1}{\alpha }\sum_{n=-\infty
}^{+\infty }\tilde{g}(2\pi n/\alpha ),  \label{Pois}
\end{equation}%
where $\tilde{g}(y)=\int_{-\infty }^{+\infty }dx\,e^{-iyx}g(x)$. For the
function corresponding to the series in Eq. (\ref{jrz}) one has $\tilde{g}%
(y)=\sqrt{\pi }e^{y^{2}/4-iys\mu }$. After the integration over $s$ we get
the expression%
\begin{equation}
\left\langle j^{r}\right\rangle =\frac{2e\beta ^{-1}}{(2\pi )^{p}V_{q}}\sum_{%
\mathbf{k}}\sum_{n=-\infty }^{+\infty }\frac{k_{r}}{\omega _{\mathbf{k}%
}^{2}+(2\pi n/\beta +i\mu )^{2}}.  \label{jrz1}
\end{equation}

The current density defined by Eq. (\ref{jrz1}) can be written as%
\begin{equation}
\left\langle j^{r}\right\rangle =\frac{2e}{L_{r}^{2}}\sum_{n_{r}=-\infty
}^{\infty }\left( 2\pi n_{r}+\tilde{\alpha}_{r}\right) \zeta _{r}(s)|_{s=1},
\label{jrz2}
\end{equation}%
with the partial zeta function
\begin{equation}
\zeta _{r}(s)=\frac{L_{r}}{\beta V_{q}}\int \frac{d\mathbf{k}_{p}}{(2\pi
)^{p}}\sum_{\mathbf{n}_{q}^{r}}\left[ \mathbf{k}_{p}^{2}+\mathbf{k}%
_{q}^{2}+\left( \frac{2\pi n_{D+1}}{\beta }+i\mu \right) ^{2}+m^{2}\right]
^{-s}.  \label{zetas}
\end{equation}%
where $\mathbf{n}_{q}^{r}=(n_{p+1},\ldots ,n_{r-1},n_{r+1},\ldots ,n_{D+1})$
and $\mathbf{k}_{q}^{2}$ is given by Eq. (\ref{omk}). Hence, in order to
find the renormalized value for the current density we need to have the
analytic continuation of the zeta function (\ref{zetas}) at the point $s=1$.

The analytic continuation can be done in a way similar to that we have used
in Ref. \cite{Bell10} for the zero temperature fermionic current. We first
integrate over the momentum along the uncompactified dimensions:%
\begin{equation}
\zeta _{r}(s)=\frac{\Gamma (s-p/2)L_{r}}{(4\pi )^{p/2}\Gamma (s)V_{q}\beta }%
\sum_{\mathbf{n}_{q}^{r}}\left[ \mathbf{k}_{q}^{2}+\left( \frac{2\pi n_{D+1}%
}{\beta }+i\mu \right) ^{2}+m^{2}\right] ^{\frac{p}{2}-s}.  \label{zeta1}
\end{equation}%
Next, the direct application of the generalized Chowla-Selberg formula \cite%
{Eliz98} to the series in Eq. (\ref{zeta1}) leads to the following expression%
\begin{eqnarray}
\zeta _{r}(s) &=&\frac{m_{r}^{D-2s}}{(4\pi )^{D/2}}\frac{\Gamma (s-D/2)}{%
\Gamma (s)}+\frac{2^{1-s}m_{r}^{D-2s}}{(2\pi )^{D/2}\Gamma (s)}  \notag \\
&&\times \sideset{}{'}{\sum}_{\mathbf{n}_{q}^{r}}\cos (\mathbf{n}%
_{q}^{r}\cdot \tilde{\boldsymbol{\alpha }}_{q})f_{\frac{D}{2}-s}(m_{r}g_{%
\mathbf{n}_{q}^{r}}(\mathbf{L}_{q}^{r})),  \label{zetadec}
\end{eqnarray}%
where $\mathbf{L}_{q}^{r}=(L_{p+1},\ldots ,L_{r-1},L_{r+1},\ldots L_{D+1})$,
$\tilde{\boldsymbol{\alpha }}_{q}=(\tilde{\alpha}_{p+1},\ldots ,\tilde{\alpha%
}_{r-1},\tilde{\alpha}_{r+1},\ldots \tilde{\alpha}_{D+1})$, with
\begin{equation}
L_{D+1}=\beta ,\;\tilde{\alpha}_{D+1}=i\mu \beta ,  \label{LD+1}
\end{equation}%
and
\begin{equation}
m_{r}^{2}=\left( 2\pi n_{r}+\tilde{\alpha}_{r}\right) ^{2}/L_{r}^{2}+m^{2}.
\label{mr2}
\end{equation}%
The prime on the summation sign in Eq. (\ref{zetadec}) means that the term $%
\mathbf{n}_{q}^{r}=0$ should be excluded from the sum and we use the notation%
\begin{equation}
g_{\mathbf{c}}(\mathbf{b})=\left(
\sum\nolimits_{i=1}^{l}c_{i}^{2}b_{i}^{2}\right) ^{1/2},  \label{gab}
\end{equation}%
for the vectors $\mathbf{c}=(c_{1},\ldots ,c_{l})$ and $\mathbf{b}%
=(b_{1},\ldots ,b_{l})$. Note that in Eq. (\ref{zetadec}), $\cos (\mathbf{n}%
_{q}^{r}\cdot \boldsymbol{\alpha }_{q})$ can also be written as $\cosh
(n_{D+1}\mu \beta )\prod\nolimits_{l=p+1,\neq r}^{D}\cos (n_{l}\tilde{\alpha}%
_{l})$.

The contribution of the second term on the right-hand side of Eq. (\ref%
{zetadec}) to the current density is finite at the physical point. The
analytic continuation is required for the part with the first term only.
That is done, by applying the summation formula (\ref{AbelPlan1}) to the
series over $n_{r}$. The further transformations are similar to that we have
used in deriving Eq. (\ref{WF4}) and we get
\begin{eqnarray}
&&\frac{\Gamma (s-D/2)}{(4\pi )^{D/2}\Gamma (s)}\sum_{n_{r}=-\infty
}^{+\infty }\frac{2\pi n_{r}+\tilde{\alpha}_{r}}{L_{r}m_{r}^{2s-D}}  \notag
\\
&&\quad =\frac{2^{2-s}m^{D+3-2s}L_{r}^{2}}{(2\pi )^{(D+1)/2}\Gamma (s)}%
\sum_{n=1}^{\infty }n\sin (n\tilde{\alpha}_{r})f_{\frac{D+3}{2}-s}(nL_{r}m).
\label{Rel2}
\end{eqnarray}%
The right-hand side of Eq. (\ref{Rel2}) is finite at the point $s=1$. Now,
substituting Eq. (\ref{zetadec}) into Eq. (\ref{jrz2}) and using Eq. (\ref%
{Rel2}), we find the following expression for the current density
\begin{eqnarray}
\left\langle j^{r}\right\rangle &=&\frac{4em^{D+1}L_{r}}{(2\pi )^{\frac{D+1}{%
2}}}\sum_{n=1}^{\infty }n\sin (n\tilde{\alpha}_{r})f_{\frac{D+1}{2}}(nL_{r}m)
\notag \\
&&+\frac{2m_{r}^{D-2}}{(2\pi )^{\frac{D}{2}}L_{r}^{2}}\sum_{n_{r}=-\infty
}^{\infty }\left( \tilde{\alpha}_{r}+2\pi n_{r}\right)  \notag \\
&&\times \sideset{}{'}{\sum}_{\mathbf{n}_{q}^{r}}\cos (\mathbf{n}%
_{q}^{r}\cdot \tilde{\boldsymbol{\alpha }}_{q})f_{\frac{D}{2}-1}(m_{r}g_{%
\mathbf{n}_{q}^{r}}(\mathbf{L}_{q}^{r})).  \label{jrz3}
\end{eqnarray}%
Note that in the limit $T\rightarrow 0$ and $L_{l}\rightarrow \infty $, $%
l\neq r$, the second term in the right-hand side of this formula vanishes.
The first term presents the current density at zero temperature in the model
with a single compact dimension (see Eq. (\ref{jrT0}) for a special case $%
p=D-1$).

An alternative representation for the expectation value of the current
density is obtained if we apply the formula (\ref{Rel3n}) to the series over
$n_{r}$ in Eq. (\ref{jrz3}). Under the condition $|\mu |\leqslant m$, this
leads to the following expression%
\begin{eqnarray}
\left\langle j^{r}\right\rangle &=&\frac{4eL_{r}m^{D+1}}{(2\pi )^{(D+1)/2}}%
\sum_{n_{r}=1}^{\infty }n_{r}\sin (n_{r}\tilde{\alpha}_{r})\sum_{\mathbf{n}%
_{q}^{r}}\cosh (n_{D+1}\mu \beta )  \notag \\
&&\times \cos (\mathbf{n}_{q-1}^{r}\cdot \tilde{\boldsymbol{\alpha }}%
_{q-1}^{r})f_{\frac{D+1}{2}}(m\sqrt{g_{\mathbf{n}_{q}}^{2}(\mathbf{L}%
_{q})+n_{D+1}^{2}\beta ^{2}}),  \label{jrz4}
\end{eqnarray}%
where $\tilde{\boldsymbol{\alpha }}_{q-1}^{r}=(\tilde{\alpha}_{p+1},\ldots ,%
\tilde{\alpha}_{r-1},\tilde{\alpha}_{r+1},\ldots \tilde{\alpha}_{D})$, $%
\mathbf{L}_{q}=(L_{p+1},\ldots ,L_{D})$, and $g_{\mathbf{n}_{q}}^{2}(\mathbf{%
L}_{q})$ is defined by Eq. (\ref{gab}). In particular, for a massless field
and for zero chemical potential, $\mu =0$, from (\ref{jrz4}) we get%
\begin{eqnarray}
\left\langle j^{r}\right\rangle &=&2eL_{r}\frac{\Gamma ((D+1)/2)}{\pi
^{(D+1)/2}}\sum_{n_{r}=1}^{\infty }n_{r}\sin (n_{r}\tilde{\alpha}_{r})
\notag \\
&&\times \sum_{\mathbf{n}_{q}^{r}}\frac{\cos (\mathbf{n}_{q-1}^{r}\cdot
\tilde{\boldsymbol{\alpha }}_{q-1}^{r})}{[g_{\mathbf{n}_{q}}^{2}(\mathbf{L}%
_{q})+n^{2}\beta ^{2}]^{(D+1)/2}}.  \label{jrz4m0}
\end{eqnarray}

The equivalence of two representations for the current density, Eqs. (\ref%
{jr}) and (\ref{jrz4}), can be seen by using the relation%
\begin{equation}
\sum_{\mathbf{n}}\cos (\mathbf{n}\cdot \boldsymbol{\alpha })f_{\nu }(c\sqrt{%
b^{2}+\sum\nolimits_{i=1}^{l}a_{i}^{2}n_{i}^{2}})=\frac{(2\pi )^{l/2}}{%
a_{1}\cdots a_{l}c^{2\nu }}\sum_{\mathbf{n}}w_{\mathbf{n}}^{2\nu -l}f_{\nu
-l/2}(bw_{\mathbf{n}}),  \label{Rel4}
\end{equation}%
where $\mathbf{n}=(n_{1},\ldots ,n_{l})$, $\boldsymbol{\alpha
}=(\alpha _{1},\ldots ,\alpha _{l})$, and $w_{\mathbf{n}}^{2}=\sum%
\nolimits_{i=1}^{l}(2\pi n_{i}+\alpha _{i})^{2}/a_{i}^{2}+c^{2}$. This
relation has been proved in Ref. \cite{Bell09} by using the Poisson's
resummation formula. Note that the formula (\ref{Rel3n}) is a special case
of Eq. (\ref{Rel4}).

An expression for the current density, convenient for the discussion of the
high-temperature limit, is obtained from Eq. (\ref{jrz4}), by applying to
the series over $n_{D+1}$ the formula (\ref{Rel3n}) under the assumption $%
|\mu |\leqslant m$. This leads to the following expression%
\begin{eqnarray}
\left\langle j^{r}\right\rangle &=&\frac{4eL_{r}}{(2\pi )^{D/2}\beta }%
\sum_{n_{r}=1}^{\infty }n_{r}\sin (n_{r}\tilde{\alpha}_{r})\sum_{\mathbf{n}%
_{q}^{r}}\cos (\mathbf{n}_{q-1}^{r}\cdot \tilde{\boldsymbol{\alpha }}_{q-1})
\notag \\
&&\times \lbrack (2\pi n_{D+1}/\beta +i\mu )^{2}+m^{2}]^{D/2}f_{D/2}(g_{%
\mathbf{n}_{q}}(\mathbf{L}_{q})\sqrt{(2\pi n_{D+1}/\beta +i\mu )^{2}+m^{2}}).
\label{jrz5}
\end{eqnarray}%
At high temperatures the dominant contribution comes from $n_{D+1}=0$ term
and to the leading order we have%
\begin{eqnarray}
\left\langle j^{r}\right\rangle &\approx &\frac{4eL_{r}T}{(2\pi )^{D/2}}%
\sum_{n_{r}=1}^{\infty }n_{r}\sin (n_{r}\tilde{\alpha}_{r})\sum_{\mathbf{n}%
_{q-1}^{r}}\cos (\mathbf{n}_{q-1}^{r}\cdot \tilde{\boldsymbol{\alpha }}%
_{q-1})  \notag \\
&&\times \left( m^{2}-\mu ^{2}\right) ^{D/2}f_{D/2}(g_{\mathbf{n}_{q}}(%
\mathbf{L}_{q})\sqrt{m^{2}-\mu ^{2}}).  \label{jrhigh}
\end{eqnarray}%
The corrections to this leading term are exponentially small. The
equivalence of two representations, Eqs. (\ref{jras}) and (\ref{jrhigh}),
for the leading order term can be seen by using the relation (\ref{Rel4}).

\subsection{Charge density}

Now we turn to the evaluation of the charge density by using the zeta
function approach. Similar to the case of Eq. (\ref{jrmode}), we have the
following mode sum%
\begin{equation}
\left\langle j_{0}\right\rangle =\frac{e}{(2\pi )^{p}V_{q}}\sum_{\mathbf{k}%
}\sum_{s=\pm }\frac{s}{e^{\beta \left( \omega _{\mathbf{k}}-s\mu \right) }-1}%
.  \label{j0mode}
\end{equation}%
The zero temperature part in the charge density vanishes due to the
cancellation between the contributions from the virtual particles and
antiparticles. The corresponding contributions to the finite temperature
part have opposite signs due to the opposite signs of the charge for
particles and antiparticles. Introducing the expectation values for the
numbers of the particles and antiparticles (per unit volume of the
uncompactified subspace),%
\begin{equation}
\left\langle N_{\pm }\right\rangle =\frac{1}{(2\pi )^{p}}\sum_{\mathbf{k}}%
\frac{1}{e^{\beta \left( \omega _{\mathbf{k}}\mp \mu \right) }-1},
\label{Npm}
\end{equation}%
the charge density is written as $\left\langle j_{0}\right\rangle
=e\left\langle N_{+}-N_{-}\right\rangle /V_{q}$. In Eq. (\ref{Npm}), the
upper/lower sign corresponds to particles/antiparticles. Note that in the
current density the contributions from particles and antiparticles have the
same sign (see Eq. (\ref{jrmode})). This is due to the fact that, though the
charges have opposite signs, the opposite signs have the velocities as well,
$v_{r}^{(+)}=k_{r}/\omega $ for particles and $v_{r}^{(-)}=-k_{r}/\omega $
for antiparticles (see the phases in the expression (\ref{fisigma}) for the
mode functions). The expression for $\left\langle N_{\pm }\right\rangle $ is
obtained from Eq. (\ref{j01}) by the replacement $2e\sinh (n\mu \beta
)/V_{q}\rightarrow e^{\pm n\mu \beta }$.

The expression (\ref{j0mode}) for the charge density may be written in the
form
\begin{equation}
\left\langle j_{0}\right\rangle =\frac{2e}{(2\pi )^{p}V_{q}}\sum_{\mathbf{k}%
}\sum_{n=1}^{\infty }e^{-n\beta \omega _{\mathbf{k}}}\sinh (n\beta \mu ).
\label{j0z}
\end{equation}%
For the further transformation of this expression we use the relation%
\begin{equation}
\,\frac{\sin (n\beta \mu )}{e^{n\beta \omega }}=\left( \mu -\int_{0}^{\mu
}d\mu \,\partial _{\beta }\beta \right) \frac{e^{-n\beta \omega }}{\omega }%
\cosh (n\beta \mu ).  \label{relsin}
\end{equation}%
As a result, the expectation value of the charge density is presented in the
form%
\begin{equation}
\left\langle j_{0}\right\rangle =\frac{2e}{(2\pi )^{p}V_{q}}\left( \mu
-\int_{0}^{\mu }d\mu \,\partial _{\beta }\beta \right) \sum_{\mathbf{k}%
}\sum_{n=1}^{\infty }\frac{e^{-n\beta \omega _{\mathbf{k}}}}{\omega _{%
\mathbf{k}}}\cosh (n\beta \mu ).  \label{j0z1}
\end{equation}

Substituting Eq. (\ref{IntRep1}), by the transformations similar to that we
have used in the case of the current density, one finds%
\begin{equation}
\left\langle j_{0}\right\rangle =2e\left. \left( \mu -\int_{0}^{\mu }d\mu
\,\partial _{\beta }\beta \right) \zeta (s)\right\vert _{s=1},  \label{j0z2}
\end{equation}%
where the corresponding zeta function is defined as%
\begin{equation}
\zeta (s)=\frac{1}{V_{q}\beta }\int \frac{d\mathbf{k}_{p}}{(2\pi )^{p}}\sum_{%
\mathbf{n}_{q}}\sum_{n=-\infty }^{+\infty }\left[ \omega _{\mathbf{k}%
}^{2}+\left( \frac{2\pi n}{\beta }+i\mu \right) ^{2}\right] ^{-s}.
\label{zeta}
\end{equation}%
with $\omega _{\mathbf{k}}$ defined by Eq. (\ref{omk}). After the
integration over over the momentum along uncompact dimensions, the function (%
\ref{zeta}) is written in the form%
\begin{equation}
\zeta (s)=\frac{\Gamma (s-p/2)}{(4\pi )^{\frac{p}{2}}\Gamma (s)V_{q}\beta }%
\sum_{\mathbf{n}_{q+1}}\left[ \sum_{l=p+1}^{D+1}\left( \frac{2\pi n_{l}+%
\tilde{\alpha}_{l}}{L_{l}}\right) ^{2}+m^{2}\right] ^{\frac{p}{2}-s},
\label{Zetaj0}
\end{equation}%
where $\mathbf{n}_{q+1}=(n_{p+1},\ldots ,n_{D+1})$ and $L_{D+1}$, $\tilde{%
\alpha}_{D+1}$ are defined by Eq. (\ref{LD+1}). The application of the
generalized Chowla-Selberg formula \cite{Eliz98} to Eq. (\ref{Zetaj0}) gives
\begin{eqnarray}
\zeta (s) &=&m^{D+1-2s}\frac{\Gamma (s-(D+1)/2)}{(4\pi )^{(D+1)/2}\Gamma (s)}%
+\frac{2^{1-s}m^{D+1-2s}}{(2\pi )^{(D+1)/2}\Gamma (s)}  \notag \\
&&\times \sideset{}{'}{\sum}_{\mathbf{n}_{q+1}}\cos (\mathbf{n}_{q+1}\cdot
\tilde{\boldsymbol{\alpha }}_{q+1})f_{\frac{D+1}{2}-s}\left( m\sqrt{g_{%
\mathbf{n}_{q}}^{2}(\mathbf{L}_{q})+n_{D+1}^{2}\beta ^{2}}\right) ,
\label{zeta2}
\end{eqnarray}%
with $\mathbf{L}_{q+1}=(L_{p+1},\ldots ,L_{D+1})$ and $\tilde{%
\boldsymbol{\alpha }}_{q+1}=(\tilde{\alpha}_{p+1},\ldots ,\tilde{\alpha}%
_{D+1})$. The prime on the summation sign in Eq. (\ref{zeta2}) means that
the term with $n_{l}=0$, $l=p+1,\ldots ,D+1$, should be excluded from the
sum.

Substituting Eq. (\ref{zeta2}) into Eq. (\ref{j0z2}), for the charge density
one finds the expression%
\begin{eqnarray}
\left\langle j_{0}\right\rangle &=&\frac{4em^{D+1}\beta }{(2\pi )^{(D+1)/2}}%
\sum_{n=1}^{\infty }n\sinh (\mu \beta n)  \notag \\
&&\times \sum_{\mathbf{n}_{q}}\cos (\mathbf{n}_{q}\cdot \tilde{%
\boldsymbol{\alpha }}_{q})f_{\frac{D+1}{2}}\left( m\sqrt{g_{\mathbf{n}%
_{q}}^{2}(\mathbf{L}_{q})+n^{2}\beta ^{2}}\right) .  \label{j0z3}
\end{eqnarray}%
Note that the first term in the right-hand side of Eq. (\ref{zeta2}) does
not depend on temperature and the corresponding contribution in Eq. (\ref%
{j0z1}) vanishes. This expression for the charge density is valid for the
region $|\mu |\leqslant m$. The equivalence of the representations (\ref{j01}%
) and (\ref{j0z3}) in this region is proved by using the formula (\ref{Rel4}%
). In Eq. (\ref{j0z3}), the term with $n_{l}=0$, $l=p+1,\ldots ,D$,
coincides with the corresponding charge density in Minkowski spacetime ($p=D$%
, $q=0$) given by Eq. (\ref{j01M}). Note that by the replacement $2e\sinh
(n\mu \beta )/V_{q}\rightarrow e^{\pm n\mu \beta }$ in Eq. (\ref{j0z3}), we
can obtain the corresponding formula for $\left\langle N_{\pm }\right\rangle
$.

An alternative expression for the charge density, convenient for the
investigation of the high-temperature limit, is obtained from Eq. (\ref{j0z3}%
) if we first separate the part corresponding to $\left\langle
j_{0}\right\rangle _{\mathrm{(M)}}$ and then apply to the series over $n$ in
the remained part the formula (\ref{Rel3n}). This leads to the following
expression:%
\begin{eqnarray}
\left\langle j_{0}\right\rangle &=&\left\langle j_{0}\right\rangle _{\mathrm{%
(M)}}-\frac{2ieT}{(2\pi )^{D/2}}\sum_{\mathbf{n}_{q}\neq 0}\cos (\mathbf{n}%
_{q}\cdot \tilde{\boldsymbol{\alpha }}_{q})\sum_{n=-\infty }^{+\infty }(2\pi
nT+i\mu )  \notag \\
&&\times \lbrack (2\pi nT+i\mu )^{2}+m^{2}]^{\frac{D}{2}-1}f_{\frac{D}{2}%
-1}(g_{\mathbf{n}_{q}}(\mathbf{L}_{q})\sqrt{(2\pi nT+i\mu )^{2}+m^{2}}).
\label{j0z4}
\end{eqnarray}%
As before, the prime means that the term with $n_{l}=0$, $l=p+1,\ldots ,D$,
should be excluded from the sum. At high temperatures the dominant
contribution to the second term in the right-hand side comes from the term
with $n=0$:%
\begin{eqnarray}
\left\langle j_{0}\right\rangle &\approx &\left\langle j_{0}\right\rangle _{%
\mathrm{M}}+\frac{2e\mu T}{(2\pi )^{D/2}}\sideset{}{'}{\sum}_{\mathbf{n}%
_{q}}\cos (\mathbf{n}_{q}\cdot \tilde{\boldsymbol{\alpha }}_{q})  \notag \\
&&\times \left( m^{2}-\mu ^{2}\right) ^{\frac{D}{2}-1}f_{\frac{D}{2}-1}(g_{%
\mathbf{n}_{q}}(\mathbf{L}_{q})\sqrt{m^{2}-\mu ^{2}}).  \label{j0high}
\end{eqnarray}%
The higher order corrections to this asymptotic expression are exponentially
small. Hence, similar to the case of the current density, the topological
part of the charge density is a linear function of the temperature in the
high-temperature limit.

In order to find the asymptotic expression for the part $\left\langle
j_{0}\right\rangle _{\mathrm{(M)}}$ at high temperatures, we use the
integral representation%
\begin{equation}
f_{\nu }(z)=\frac{2^{-\nu }\sqrt{\pi }}{\Gamma (\nu +1/2)}\int_{1}^{\infty
}dt\,(t^{2}-1)^{\nu -1/2}e^{-zt}.  \label{fnuint}
\end{equation}%
Substituting this into Eq. (\ref{j01M}) and changing the orders of
integration and summation, the summation is done explicitly and to the
leading order we get%
\begin{equation}
\left\langle j_{0}\right\rangle _{\mathrm{(M)}}\approx 2e\mu T^{D-1}\frac{%
\Gamma ((D+1)/2)}{\pi ^{(D+1)/2}}\zeta _{\mathrm{R}}(D-1),  \label{j0Mhigh}
\end{equation}%
where $\zeta _{\mathrm{R}}(x)$ is the Riemann zeta function. This result has
been obtained in Ref. \cite{Habe81}. As it is seen from Eq. (\ref{j0high}),
for $D>2$ the Minkowskian part dominates in the high temperature limit and
one has $\left\langle j_{0}\right\rangle \approx \left\langle
j_{0}\right\rangle _{\mathrm{(M)}}$.

\section{Bose-Einstein condensation}

\label{Sec:BEC}

In this section we consider the application of the formulas given before for
the investigation of the Bose-Einstein condensation (BEC). This phenomena
for a relativistic Bose gas of scalar particles in topologically trivial
flat spacetime has been discussed in Refs. \cite{Kapu81,Habe81,Bern91}. The
investigation of the critical behavior of an ideal Bose gas confined to the
background geometry of a static Einstein universe is given in Refs. \cite%
{Sing84} for scalar and vector fields. BEC in higher dimensional spacetime
with $S^{N}$ as a compact subspace has been considered \ in Ref. \cite%
{Shir87}. The case of an ultrastatic (3+1)-dimensional manifold with a
hyperbolic spatial part is analyzed in Ref. \cite{Cogn93}. The background
geometry of closed Robertson-Walker spacetime is discussed in Ref. \cite%
{Truc98}. Thermodynamics of ideal boson and fermion gases in anti-de Sitter
spacetime and in the static Taub universe have been considered in Refs. \cite%
{Vanz94,Huan94}. In the high-temperature limit, BEC in a general background
has been discussed in Refs. \cite{Toms92,Kirs95,Smit96}. Recently, BEC on
product manifolds, when the gas of bosons is confined by anisotropic
harmonic oscillator potential, has been investigated in Ref. \cite{Fucc11}.

Note that in the literature two criteria have been considered for BEC (see,
for instance, the discussion in Refs. \cite{Kirs96}). In the first one the
existence of critical temperature $T_{c}>0$ is assumed for which the
chemical potential becomes equal to the single particle ground state energy.
The derivative $\partial _{T}\mu $ for a fixed value of a conserved charge
is discontinuous at the critical temperature and the condensation
corresponds to a phase transition. By the second criterion, one assumes the
existence of a finite fraction of particle density in the ground state and
in states in its neighborhood at $T>0$. In this case the presence of a phase
transition is not required and thermodynamical functions can be continuous.
In particular, in Ref. \cite{Habe81} it has been shown that for massive
particles there is no BEC in dimensions $D\leqslant 2$ if one follows the
first criterion.

In the discussion of previous sections we have considered the charge and
current densities as functions of the temperature, chemical potential and
the lengths of compact dimensions. From the physical point of view it is
more important to consider the behavior of the system for a fixed value of
the charge. We will denote by $Q$ the charge per unit volume of the
uncompactified subspace, $Q=V_{q}\left\langle j_{0}\right\rangle $. From Eq.
(\ref{j01}) for this quantity one has
\begin{equation}
Q=\frac{4e\beta }{(2\pi )^{\frac{p+1}{2}}}\sum_{n=1}^{\infty }n\sinh (n\mu
\beta )\sum_{\mathbf{n}_{q}}\omega _{\mathbf{n}_{q}}^{p+1}f_{\frac{p+1}{2}%
}(n\beta \omega _{\mathbf{n}_{q}}).  \label{q}
\end{equation}%
For a fixed value of the charge, this relation implicitly determines the
chemical potential as a function of the temperature, lengths of the compact
dimensions and the charge.

For high temperatures the chemical potential determined from Eq. (\ref{q})
tends to zero. Hence, at high temperatures we always have solution with $%
|\mu |<\omega _{0}$. The further behavior of the function $\mu (T)$ with
decreasing temperature is essentially different in the cases $p>2$ and $%
p\leqslant 2$. For $p>2$ the expression (\ref{q}) is finite in the limit $%
|\mu |\rightarrow \omega _{0}$. We denote by $T_{c}$ the temperature at
which one has $|\mu (T_{c})|=\omega _{0}$ for a fixed value of the charge.
It is the critical temperature for BEC. The formula
\begin{equation}
|Q|=\frac{4|e|\beta _{c}}{(2\pi )^{\frac{p+1}{2}}}\sum_{n=1}^{\infty }n\sinh
(n\omega _{0}\beta _{c})\sum_{\mathbf{n}_{q}}\omega _{\mathbf{n}%
_{q}}^{p+1}f_{\frac{p+1}{2}}(n\beta _{c}\omega _{\mathbf{n}_{q}}),
\label{Tc}
\end{equation}%
with $\beta _{c}=1/T_{c}$, determines the critical temperature as a function
of the charge, of the lengths of the compact dimensions, and of the
parameters $\tilde{\alpha}_{l}$.

\begin{figure}[tbph]
\begin{center}
\epsfig{figure=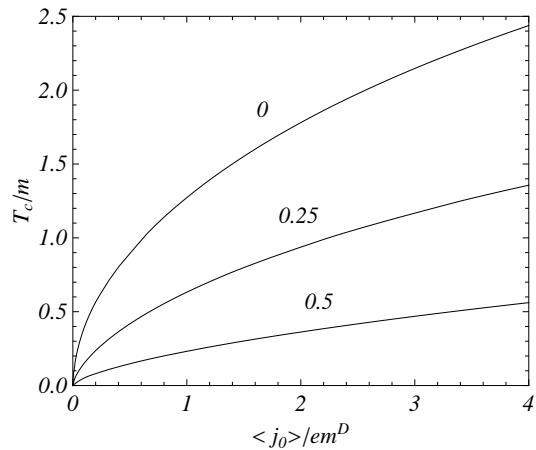,width=7.cm,height=6.cm}
\end{center}
\caption{The critical temperature as a function of the charge density in the
$D=4$ model with a single compact dimension. The graphs are plotted for $%
L_{D}m=0.5$ and for different values of $\tilde{\protect\alpha}_{D}$ and the
numbers near the curves correspond to the values of the parameter $\tilde{%
\protect\alpha}_{D}/(2\protect\pi )$.}
\label{fig2}
\end{figure}

Simple asymptotic formulas for the critical temperature are obtained for low
and high temperatures. At low temperatures, $\omega _{0}\beta _{c}\gg 1$,
the dominant contribution in Eq. (\ref{Tc}) comes from the mode with the
smallest energy corresponding to $n_{l}=0$. By using the asymptotic
expression for the Macdonald function for large values of the argument, from
Eq. (\ref{Tc}) to the leading order one finds
\begin{equation}
T_{c}\approx \frac{2\pi }{\omega _{0}}\left[ \frac{|Q/e|}{\zeta _{\mathrm{R}%
}(p/2)}\right] ^{2/p}.  \label{Tclow}
\end{equation}%
This regime is realized for values of the charge corresponding to $|Q/e|\ll
\omega _{0}^{p}$. At high temperatures, by taking into account that to the
leading order $\left\langle j_{0}\right\rangle \approx \left\langle
j_{0}\right\rangle _{\mathrm{(M)}}$ and by using the asymptotic expression (%
\ref{j0Mhigh}), one finds%
\begin{equation}
T_{c}\approx \left[ \frac{\pi ^{(D+1)/2}|Q/e|V_{q}^{-1}}{2\omega _{0}\Gamma
((D+1)/2)\zeta _{R}(D-1)}\right] ^{1/(D-1)}.  \label{TcHigh}
\end{equation}%
This asymptotic corresponds to $|Q/e|\gg V_{q}\omega _{0}^{D}$. In figure %
\ref{fig2} we display the critical temperature as a function of the charge
density in the $D=4$ model with a single compact dimension ($p=3$, $q=1$).
The graphs are plotted for $mL_{D}=0.5$ and for different values of $\tilde{%
\alpha}_{D}/(2\pi )$ (numbers near the curves). As it is seen, for fixed
lengths of the compact dimensions, the critical temperature for the phase
transition can be controlled by tuning the value for the gauge potential.

\begin{figure}[tbph]
\begin{center}
\begin{tabular}{cc}
\epsfig{figure=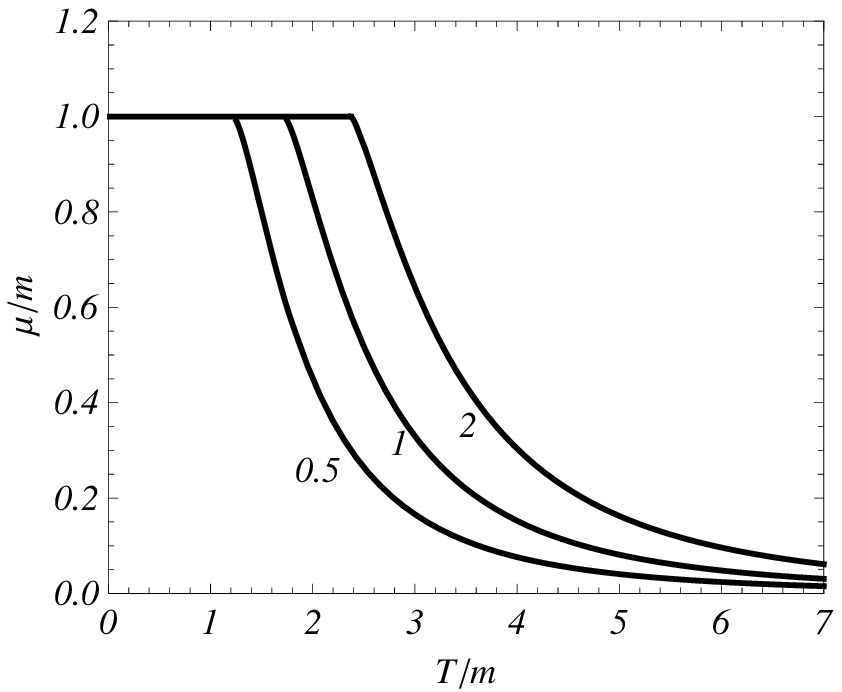,width=7.cm,height=6.cm} & \quad %
\epsfig{figure=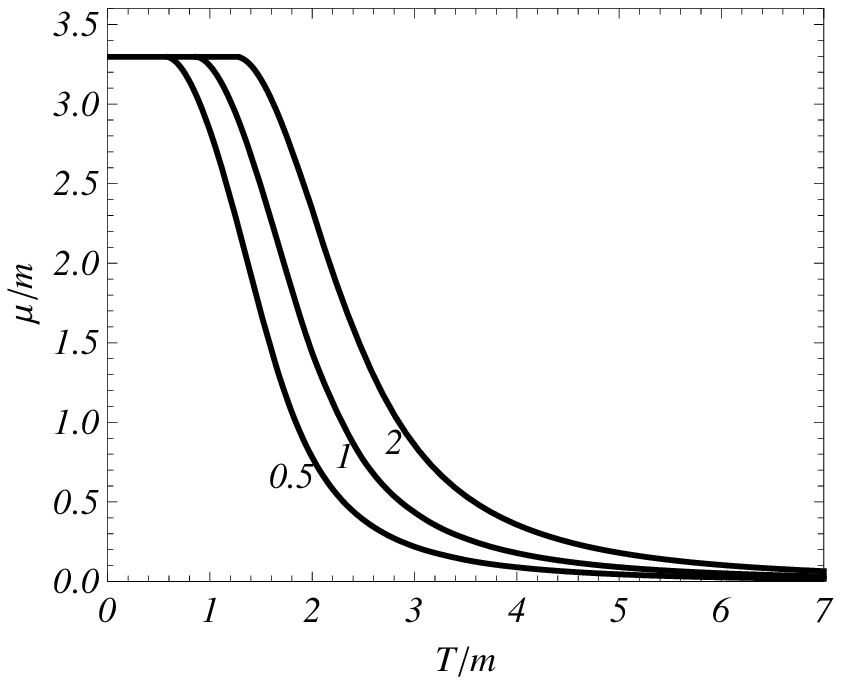,width=7.cm,height=6.cm}%
\end{tabular}%
\end{center}
\caption{The chemical potential as a function of the temperature in the $D=4$
model with a single compact dimension for $mL_{D}=0.5$. The left and right
plots correspond to $\tilde{\protect\alpha}_{D}=0$ and $\tilde{\protect\alpha%
}_{D}=\protect\pi /2$ respectively. The numbers near the curves correspond
to the values of the parameter $m^{1-D}Q/e$. }
\label{fig3}
\end{figure}

At temperatures $T<T_{c}$, the equation (\ref{q}) has no solutions with $%
|\mu |<\omega _{0}$. The consideration in this region of temperature is
similar to the standard one for the BEC in topologically trivial spaces. We
note that the expression (\ref{q}) does not include the charge corresponding
to the states with $\mathbf{k}_{p}=0$. At temperatures $T<T_{c}$ the
expression (\ref{q}) with $|\mu |=\omega _{0}$ determines the charge
corresponding to the states with $\mathbf{k}_{p}\neq 0$. We denote this
charge by $Q_{1}$:%
\begin{equation}
Q_{1}=\frac{4e\beta \mathrm{sgn}(\mu )}{(2\pi )^{\frac{p+1}{2}}}%
\sum_{n=1}^{\infty }n\sinh (n\omega _{0}\beta )\sum_{\mathbf{n}_{q}}\omega _{%
\mathbf{n}_{q}}^{p+1}f_{\frac{p+1}{2}}(n\beta \omega _{\mathbf{n}_{q}}).
\label{q1}
\end{equation}%
For the charge corresponding to the Bose-Einstein condensate at the state $%
\mathbf{k}_{p}=0$ one has $Q_{c}=Q-Q_{1}$. This charge vanishes at $T=T_{c}$%
. At low temperatures, by making use of Eq. (\ref{Tclow}), for the
corresponding charges below the critical temperature one finds%
\begin{equation}
Q_{1}=Q(T/T_{c})^{p/2},\;Q_{c}=Q\left[ 1-(T/T_{c})^{p/2}\right] .
\label{Qlow}
\end{equation}%
For high temperatures we use the asymptotic formula (\ref{TcHigh}) with the
results:%
\begin{equation}
Q_{1}=Q(T/T_{c})^{D-1},\;Q_{c}=Q\left[ 1-(T/T_{c})^{D-1}\right] .
\label{Qhigh}
\end{equation}%
In particular, Eq. (\ref{Qhigh}) coincides with the corresponding result in
Minkowski spacetime \cite{Habe81}. In figure \ref{fig3}, we plot the
chemical potential as a function of the temperature in the $D=4$ model with
a single compact dimension ($p=3$, $q=1$). The left and right plots
correspond to $\tilde{\alpha}_{D}=0$ and $\tilde{\alpha}_{D}=\pi /2$
respectively. For the length of the compact dimension we have taken the
value corresponding to $mL_{D}=0.5$ and the numbers near the curves
correspond to the values of the parameter $m^{1-D}Q/e$.

Note that for $T<T_{c}$ the scalar field acquires a nonzero ground-state
expectation value $\varphi _{c}$. The latter can be found in a way similar
to that used in Ref. \cite{Toms92} (see also Refs. \cite{Kapu81,Habe81} ):%
\begin{equation}
|\varphi _{c}|^{2}=\frac{|(Q-Q_{1})/e|}{2\omega _{0}V_{q}},  \label{phig}
\end{equation}%
with $Q$ and $Q_{1}$ given by Eqs. (\ref{q}) and (\ref{q1}).

Having the chemical potential from Eq. (\ref{q}) for $T>T_{c}$ and taking $%
|\mu |=\omega _{0}$ for $T<T_{c}$, we can evaluate the current density for a
fixed value of the charge by using Eq. (\ref{jr2}). Note that at high
temperatures the chemical potential tends to zero and the leading term in
the corresponding asymptotic expansion for the current density is obtained
from Eq. (\ref{jrhigh}) with $\mu =0$. For $T<T_{c}$, the formula (\ref{jr2}%
) gives a part of the current density due to the excited states only. In
addition to this, there is a contribution due to the condensate, given by
the expression%
\begin{equation}
j_{c}^{r}=\frac{\tilde{\alpha}_{r}Q_{c}}{L_{r}\omega _{0}V_{q}},  \label{jrc}
\end{equation}%
with $r=p+1,\ldots ,D$ and $|\tilde{\alpha}_{r}|\leqslant \pi $. For the
expectation value of the total current density one has $\left\langle
j^{r}\right\rangle _{t}=j_{c}^{r}+\left\langle j^{r}\right\rangle $.

\begin{figure}[tbph]
\begin{center}
\begin{tabular}{cc}
\epsfig{figure=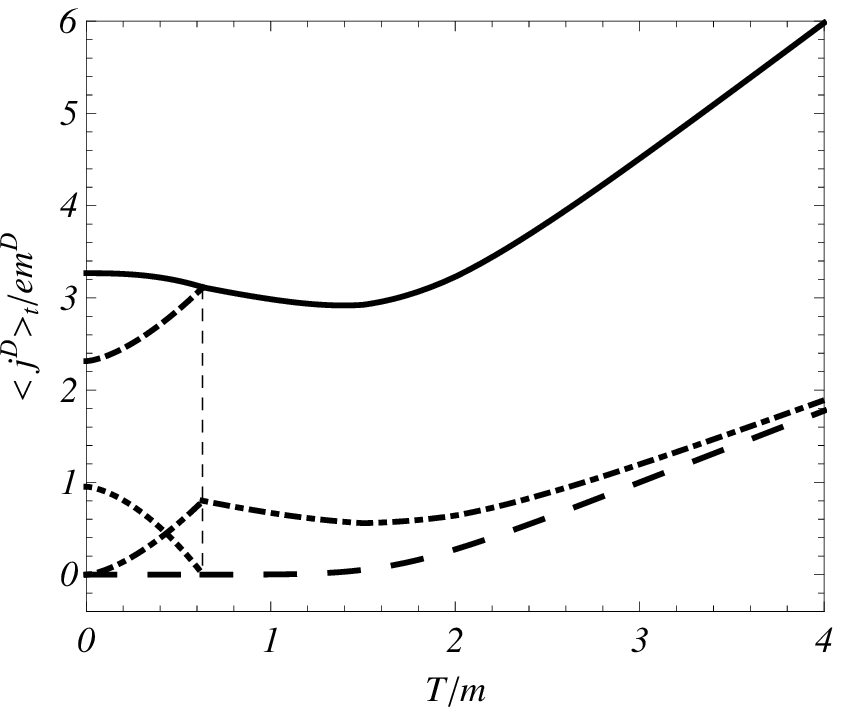,width=7.cm,height=6.cm} & \quad %
\epsfig{figure=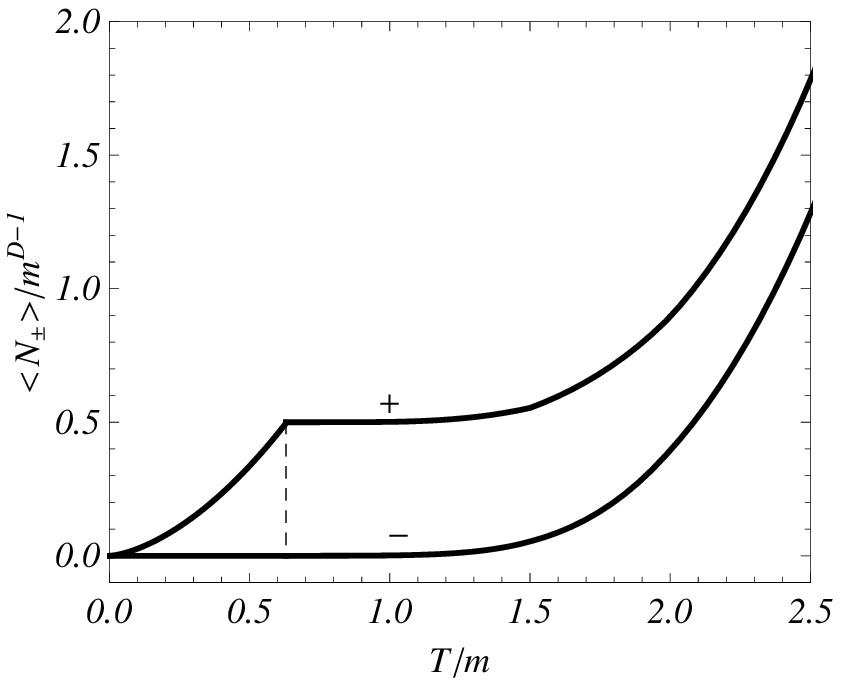,width=7.cm,height=6.cm}%
\end{tabular}%
\end{center}
\caption{The expectation values of the current density along the compact
dimension (left plot) and of the particle/antiparticle numbers (right plot)
as functions of the temperature in the $D=4$ model for $mL_{D}=0.5$, $\tilde{%
\protect\alpha}_{D}=\protect\pi /2$ and $Q/e=0.5m^{D-1}$. }
\label{fig4}
\end{figure}

In figure \ref{fig4} we plot the expectation values of the total current
density (left plot, full curve) and of the particle and antiparticle numbers
in the states with $\mathbf{k}_{p}\neq 0$, $\left\langle N_{\pm
}\right\rangle $ (right plot), versus the temperature for $mL_{D}=0.5$, $%
\tilde{\alpha}_{D}=\pi /2$ and $Q/e=0.5m^{D-1}$ in the $D=4$ model with a
single compact dimension. For the corresponding critical temperature from
Eq. (\ref{Tc}) one finds $T_{c}\approx 0.63 m$. In the left plot we have
separately presented the contributions to the current density coming from
particles (dot-dashed line) and antiparticles (large dashed line). The
linear dependence at high temperatures is clearly seen. On the left panel we
have also plotted the separate contributions to the current density from the
excited states, $\left\langle j^{D}\right\rangle /(em^{D})$ (dashed line),
and from the condensate, $j_{c}^{D}/(em^{D})$ (dotted line). Note that for
the current density given by Eq. (\ref{jrT0}) we have $\left\langle
j^{D}\right\rangle _{0}=2.32em^{D}$. The total current and its first
derivative with respect of the temperature are continuous functions at the
point of the phase transition. The latter is not the case for the separate
parts coming from the condensate and from the excited states.

\begin{figure}[tbph]
\begin{center}
\epsfig{figure=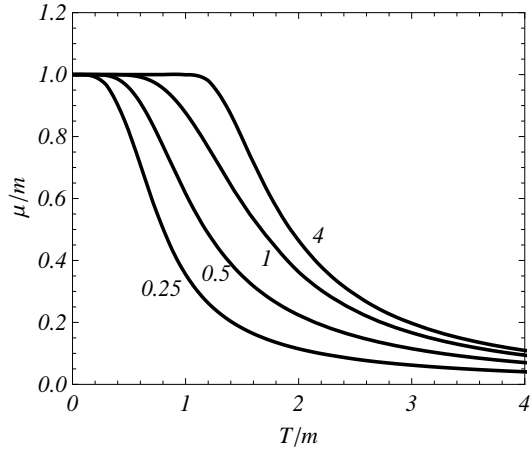,width=7.cm,height=6.cm}
\end{center}
\caption{The chemical potential as a function of the temperature in the $D=3$
model with a single compact dimension for $\tilde{\protect\alpha}_{D}=0$ and
$m^{1-D}\left\langle j_{0}\right\rangle /e=0.5$. The numbers near the curves
correspond to the values of the parameter $L_{D}m$.}
\label{fig5}
\end{figure}

For $p\leqslant 2$ the charge defined by Eq. (\ref{q}) diverges in the limit
$|\mu |\rightarrow \omega _{0}$ and for finite value of charge density the
point $|\mu |=m$ cannot be reached. The corresponding asymptotic expressions
for the charge and current densities are given by Eqs. (\ref{j0mutoom0}) and
(\ref{jrmutoom0}). Thus, for $p\leqslant 2$ there is no BEC by the first
criterion given above. In particular, this is the case in the model with
compact space corresponding to $p=0$. This result for spaces of finite
volume in general has been obtained in Ref. \cite{Smit96}. As before, for $%
p\leqslant 2$ and for a fixed value of the charge $Q$, the dependence of the
chemical potential on the temperature is determined by Eq. (\ref{q}). In the
limit $\beta \rightarrow \infty $ (low temperatures) and for a fixed value
of $|\mu |\neq \omega _{0}$, the expression on the right-hand side of Eq. (%
\ref{q}) tends to zero. From here we conclude that for a fixed value of $Q$
we should have $|\mu |\rightarrow \omega _{0}$ for $\beta \rightarrow \infty
$. The corresponding asymptotic behavior is found in a way similar to that
we have used for Eq. (\ref{muom}) and is given by the same expression.
Solving with respect to the chemical potential, we find the following
asymptotic expressions%
\begin{eqnarray}
|\mu | &\approx &\omega _{0}-\left( \frac{\omega _{0}}{2\pi }\right) ^{\frac{%
p}{2-p}}\left[ \frac{|e|}{|Q|}\Gamma \left( 1-\frac{p}{2}\right) T\right] ^{%
\frac{2}{2-p}},\;p=0,1,  \notag \\
|\mu | &\approx &\omega _{0}-T\exp \left[ -\frac{|Q|}{|e|}\left( \frac{2\pi
}{\omega _{0}T}\right) ^{p/2}\right] ,\;p=2,  \label{muNBE}
\end{eqnarray}%
in the limit $T\rightarrow 0$. In figure \ref{fig5} we have plotted the
chemical potential versus temperature in the $D=3$ model with a single
compact dimension ($p=2$, $q=1$) for $\tilde{\alpha}_{D}=0$ and for a fixed
value of the charge density corresponding to $m^{1-D}\left\langle
j_{0}\right\rangle /e=0.5$. The numbers near the curves correspond to the
values of the parameter $L_{D}m$. As it is seen, though for a finite value
of $L_{D}$ the derivative $\partial _{T}\mu (T)$ is a continuous function,
in the limit $L_{D}\rightarrow \infty $ it tends to the corresponding
function in the $D=3$ model with trivial topology ($p=3$, $q=0$) for which
the function $\partial _{T}\mu (T)$ is discontinuous at the point $T=T_{c}$.

\section{Conclusion}

\label{Sec:Conc}

In the present paper we have investigated the finite temperature expectation
values of the charge and current densities for a complex scalar field,
induced by nontrivial spatial topology. As an example for the latter we have
considered a flat spacetime with an arbitrary number of toroidally
compactified dimensions. This allowed us to escape the problems related to
the curvature and to extract pure topological effects. The periodicity
conditions along compact dimensions are taken in the form (\ref{BC_pq}) with
general constant phases. As \ special cases the latter includes the
periodicity conditions for untwisted and twisted fields. In addition, we
have assumed the presence of a constant gauge field. By performing a gauge
transformation, the gauge field is excluded from the field equation.
However, this leads to the shift in the phases appearing in the periodicity
conditions given by Eq. (\ref{alftilde}).

In the evaluation of the expectation values for the charge and current
densities we have used two different approaches which allowed us to obtain
alternative representations for the corresponding expectation values. In the
first approach we evaluated the thermal Hadamard function by using the
Abel-Plana-type summation formula for the series over the momentum along a
compact dimension. The corresponding expression is given by Eq. (\ref{WF4}).
The $n_{r}=0$ term in that formula corresponds to the Hadamard function for
the topology $R^{p+1}\times (S^{1})^{q-1}$ and, hence, the $n_{r}\neq 0$
part is the change in the Hadamard function due to the compactification of
the $r$-th direction. An alternative representation for the Hadamard
function is given by Eq. (\ref{G1alt}).

Given the Hadamard function, the expectation values of the charge and
current densities are evaluated by making use of Eq. (\ref{jW}). The charge
density is given by two equivalent representations, Eqs. (\ref{j0}) and (\ref%
{j01}). The explicit information contained in Eq. (\ref{j0}) is more
detailed. The term with $n_{r}=0$ in this representation corresponds to the
charge density for the topology $R^{p+1}\times (S^{1})^{q-1}$ with the
lengths of the compact dimensions $(L_{p+1},\ldots ,L_{r-1},L_{r+1},\ldots
,L_{D})$ and the contribution of the terms with $n_{r}\neq 0$ is the change
in the current density induced by the compactification of the $r$-th
dimension. The charge density is an even periodic function of the phases $%
\tilde{\alpha}_{l}$ with the period equal to $2\pi $. The sign of the ratio $%
\left\langle j_{0}\right\rangle /e$ coincides with the sign of the chemical
potential. If the length of the $l$-th compact dimension is small compared
with the other length scales and $L_{l}\ll \beta $, the behavior of the
charge density is essentially different with dependence whether the
parameter $\tilde{\alpha}_{l}$ is zero or not. For $\tilde{\alpha}_{l}=0$,
to the leading order, $L_{l}\left\langle j_{0}\right\rangle $ coincides with
the charge density in $(D-1)$-dimensional space with topology $R^{p}\times
(S^{1})^{q-1}$ and with the lengths of the compact dimensions $L_{p+1}$%
,\ldots ,$L_{l-1}$,$L_{l+1}$,\ldots ,$L_{D}$. For $\tilde{\alpha}_{l}\neq 0$
the charge density is suppressed by the factor $e^{-|\tilde{\alpha}%
_{l}|\beta /L_{l}}$. At low temperatures and for a fixed value of $|\mu
|<\omega _{0}$, the charge density is suppressed by the factor $e^{-(\omega
_{0}-|\mu |)/T}$. For a fixed temperature and in the limit $|\mu
|\rightarrow \omega _{0}$ the charge density is finite for $p>2$ and it
diverges for $p\leqslant 2$. The corresponding asymptotic behavior in the
latter case is given by Eq. (\ref{j0mutoom0}). In the high-temperature limit
and for $D>2$ the Minkowskian part dominates in the charge density with the
leading term given by Eq. (\ref{j0Mhigh}). In the same limit, the
topological part of the charge density is a linear function of the
temperature.

For the expectation value of the current density along the $r$-th compact
dimension we have derived representations given by Eqs. (\ref{jr}) and (\ref%
{jr2}). The components along uncompactified dimensions vanish. The current
density along the $r$-th compact dimension is an odd periodic function of $%
\tilde{\alpha}_{r}$ and an even periodic function of $\tilde{\alpha}_{l}$, $%
l\neq r$, with the period equal to $2\pi $. The current density is an even
function of the chemical potential. Unlike to the case of the charge
density, the current density does not vanish at zero temperature for a fixed
value of the chemical potential. The corresponding expression is given by
Eq. (\ref{jrT0}) and the properties are discussed in Appendix \ref%
{sec:Appendix}. For small values of $L_{l}$, $l\neq r$, and for $\tilde{%
\alpha}_{l}=0$, to the leading order, the quantity $L_{l}\left\langle
j^{r}\right\rangle $ coincides with the $r$-th component of the current
density in $(D-1)$-dimensional space with topology $R^{p}\times
(S^{1})^{q-1} $ and with the lengths of the compact dimensions $L_{p+1}$%
,\ldots ,$L_{l-1}$,$L_{l+1}$,\ldots ,$L_{D}$. For $\tilde{\alpha}_{l}\neq 0$
and for small values of $L_{l}$, $l\neq r$, the current density $%
\left\langle j^{r}\right\rangle $ is exponentially suppressed. At a fixed
temperature and for $p\leqslant 2$ the current density is divergent in the
limit $|\mu |\rightarrow \omega _{0}$. The leading term in the corresponding
asymptotic expansion is related to the charge density by Eq. (\ref{jrmutoom0}%
). For a fixed value of the chemical potential $|\mu |<\omega _{0}$ and at
low temperatures the finite temperature corrections are given by Eq. (\ref%
{jr2low}) and they are exponentially small. In the limit of high
temperatures, the current density is a linear function of the temperature.

In Section \ref{Sec:Zeta}, we have derived alternative representations for
the expectation values of the charge and current densities by using the zeta
function approach. In both cases, by applying to the corresponding zeta
functions the generalized Chowla-Selberg formula, Eqs. (\ref{jrz4}), (\ref%
{jrz5}) and Eqs. (\ref{j0z3}), (\ref{j0z4}) are obtained for the current and
charge densities respectively. At high temperatures, the leading term in the
asymptotic expansion of the current density is given by Eq. (\ref{jrhigh})
with the linear dependence on the temperature and the next corrections are
exponentially small. For the charge density, for $D>2$ the leading term in
the high-temperature expansion coincides with the corresponding charge
density in $(D+1)$-dimensional Minkowskian spacetime. The leading term in
the correction induced by nontrivial topology linearly depends on the
temperature and the following corrections are exponentially suppressed.

The Bose-Einstein condensation is discussed in section
\ref{Sec:BEC}. For a fixed value of the charge, the relation
(\ref{q}) determines the chemical potential as a function of the
temperature, of the lengths of compact directions and of the
phases in the periodicity conditions. For high temperatures the
chemical potential tends to zero. With decreasing temperature the
chemical potential increases and for $p>2$ one has $|\mu
(T)|=\omega _{0}$ at some finite temperature $T=T_{c}$. The
critical temperature for BEC, $T_{c}$, is determined by Eq.
(\ref{Tc}). Simple expressions are obtained for low and high
temperatures, Eqs. (\ref{Tclow}) and (\ref{TcHigh}), respectively.
At temperatures $T<T_{c}$ one has $|\mu |=\omega _{0}$ and Eq.
(\ref{q}) determines the charge corresponding to the states with
$\mathbf{k}_{p}\neq 0$ and the remained charge corresponds to the
charge of the condensate. At low
and high temperatures the charges are given by simple expressions (\ref{Qlow}%
) and (\ref{Qhigh}). Similar to the charge density, for $T<T_{c}$ the
current density is the sum of two parts. The first one is the contribution
of excited states and is given by Eq. (\ref{jr2}) with $|\mu |=\omega _{0}$.
The second part is due to the condensate and it is presented by Eq. (\ref%
{jrc}). The total current and its first derivative with respect to the
temperature are continuous functions at the critical temperature. For $%
p\leqslant 2$, the point $|\mu |=\omega _{0}$ cannot be reached for finite
value of charge density. For a fixed value of the charge, we have $|\mu
|\rightarrow \omega _{0}$ in the limit $T\rightarrow 0$. The corresponding
asymptotic behavior is given by Eq. (\ref{muNBE}). In this case the
thermodynamical functions are continuous and there is no phase transition at
finite temperature.

\section{Acknowledgments}

E.R.B.M. thanks Conselho Nacional de Desenvolvimento Cient\'{\i}fico e Tecnol%
\'{o}gico (CNPq) for partial financial support. A.A.S. was supported by CNPq.

\appendix

\section{Vacuum expectation value of the current density}

\label{sec:Appendix}

In this Appendix we give some properties of the zero temperature current
density given by Eq. (\ref{jrT0}). An alternative expression is obtained
from Eq. (\ref{jrz4}) taking the limit $\beta \rightarrow \infty $. In this
limit the term with $n=0$ survives only and we get:%
\begin{eqnarray}
\left\langle j^{r}\right\rangle _{0} &=&\frac{4eL_{r}m^{D+1}}{(2\pi
)^{(D+1)/2}}\sum_{n_{r}=1}^{\infty }n_{r}\sin (n_{r}\tilde{\alpha}_{r})
\notag \\
&&\times \sum_{\mathbf{n}_{q-1}^{r}}\cos (\mathbf{n}_{q-1}^{r}\cdot %
\boldsymbol{\alpha }_{q-1})f_{\frac{D+1}{2}}(mg_{\mathbf{n}_{q}}(\mathbf{L}%
_{q})).  \label{jrT0b}
\end{eqnarray}%
Let us consider the behavior of the zero temperature current density in some
limiting cases. First we consider the limit when the length of the $r$-th
compact dimension, $L_{r}$, is much larger than the other length scales. The
behavior of the current density in this limit crucially depends whether $%
\omega _{0r}$, defined by (\ref{om0r}), is zero or not. In the
first case, which is realized for $\tilde{\alpha}_{l}=0$, $l\neq
r$, and $m=0$, the dominant contribution in Eq. (\ref{jrT0}) for
large values of $L_{r}$ comes from
the modes with $n_{l}=0$, $l\neq r$, for which $\omega _{\mathbf{n}%
_{q-1}^{r}}=\omega _{0r}=0$. The corresponding expression is
obtained from Eq. (\ref{jrT0}) taking the limit $\omega
_{\mathbf{n}_{q-1}^{r}}\rightarrow 0$ and to the leading order we
have
\begin{equation}
\left\langle j^{r}\right\rangle _{0}\approx \frac{2e\Gamma (p/2+1)}{\pi
^{p/2+1}L_{r}^{p}V_{q}}\sum_{n_{r}=1}^{\infty }\frac{\sin (n_{r}\tilde{\alpha%
}_{r})}{n_{r}^{p+1}}.  \label{jrT0zero}
\end{equation}%
For $\omega _{0r}\neq 0$ and for large values of $L_{r}$, the main
contribution to the zero temperature current density comes from the mode $%
n_{r}=1$, $n_{l}=0$, $l\neq r$, and from Eq. (\ref{jrT0}) one finds%
\begin{equation}
\left\langle j^{r}\right\rangle _{0}\approx \frac{2eV_{q}^{-1}\sin (\tilde{%
\alpha}_{r})\omega _{0r}^{(p+1)/2}}{(2\pi )^{(p+1)/2}L_{r}^{(p-1)/2}}%
e^{-L_{r}\omega _{0r}}.  \label{jrT0large}
\end{equation}%
In this case we have an exponential suppression.

Now we discuss the asymptotic of the current density for small values of $%
L_{r}$. In this limit it is more convenient to use Eq. (\ref{jrT0b}). First
we separate the term $n_{l}=0$, $l\neq r$, in Eq. (\ref{jrT0b}) and use the
asymptotic expression of the Macdonald function for small values of the
argument. For the remained part in Eq. (\ref{jrT0b}), the dominant
contribution comes from large values of $n_{r}$ and, to the leading order,
we replace the summation over $n_{r}$ by the integration. The corresponding
integral involving the Macdonald function is evaluated by using the formula
from Ref. \cite{Prud86}. In this way it can be seen that the contribution of
the mode with a given $\mathbf{n}_{q-1}^{r}$ is suppressed by the factor $%
\exp (-g_{r}\sqrt{\tilde{\alpha}_{r}^{2}/L_{r}^{2}+m^{2}})$, where $%
g_{r}=\sum\nolimits_{l=p+1,\neq r}^{D}n_{l}^{2}L_{l}^{2}$. As a result, we
see that the dominant contribution to the current density is due to the
modes with $n_{l}=0$, $l\neq r$, and to the leading order we get%
\begin{equation}
\left\langle j^{r}\right\rangle _{0}\approx \frac{2e\Gamma ((D+1)/2)}{\pi
^{(D+1)/2}L_{r}^{D}}\sum_{n_{r}=1}^{\infty }\frac{\sin (n_{r}\tilde{\alpha}%
_{r})}{n_{r}^{D}}.  \label{jrT0small}
\end{equation}%
This leading term does not depend on the mass and on the lengths of the
other compact dimensions. As it is seen from Eq. (\ref{jrT0b}), the
expression in the right-hand side of Eq. (\ref{jrT0small}) coincides with
the current density for a massless scalar field in the space with topology $%
R^{D-1}\times S^{1}$.

\end{document}